\documentclass[journal]{IEEEtran}
\IEEEoverridecommandlockouts
\usepackage{times,amsmath,color,amssymb,graphicx,epsfig,cite,psfrag,balance}
\usepackage{svg}
\usepackage[caption=false,font=footnotesize]{subfig}

\usepackage{caption}
\DeclareCaptionLabelFormat{tcomm}{Fig. \thefigure.}  
\usepackage{capt-of}
\usepackage{amsfonts,pifont,enumerate,cases}
\usepackage{mathrsfs} 
\usepackage[table]{xcolor} 
\usepackage{verbatim} 
\usepackage{bm}
\usepackage{cuted,stfloats}
\usepackage{algorithm}
\usepackage{algorithmic}
\usepackage{longtable}
\usepackage{blindtext}
\usepackage{multirow}
\usepackage{float}
\usepackage{threeparttable}
\usepackage{makecell}
\usepackage[utf8]{inputenc}
\usepackage{url}
\usepackage{booktabs}
\usepackage{amssymb}
\usepackage{bbding}
\usepackage{pifont}
\usepackage{wasysym}
\usepackage{utfsym}
\usepackage{fontawesome}
\usepackage[algo2e,ruled,linesnumbered,lined,boxed,commentsnumbered]{algorithm2e}
\usepackage[
    colorlinks=true,
    linkcolor=blue,
    citecolor=blue,
    urlcolor=magenta
]{hyperref}

\begin{document}
\title{Discrete Antenna Positioning and Beamforming Design for RIS-Assisted MA Secure ISAC Systems}
\author{Zhendong Li, Mingze Zhu, Zhou Su, Lin Chen, Kang Wei, Wen Fang, Ying Wang, and Wen Chen
\thanks{Zhendong Li and Mingze Zhu are with the School of Information and Communication Engineering, Xi’an Jiaotong University, Xi’an 710049, China (email: lizhendong@xjtu.edu.cn; 2684484545@stu.xjtu.edu.cn). Zhou Su is with the School of Cyber Science and Engineering, Xi'an Jiaotong University, Xi'an 710049, China (email: zhousu@ieee.org). Lin Chen is with the Department of Electrical and Computer Engineering, Stevens Institute of Technology, Hoboken, NJ 07030, USA (e-mail: lchen53@stevens.edu). Kang Wei is with the School of Cyber Science and Engineering, Southeast University, Nanjing 211189, China (e-mail: kang.wei@seu.edu.cn). Wen Fang is with the College of Electronic and Information Engineering, Tongji University, Shanghai 201804, China (e-mail: wen.fang@tongji.edu.cn). Ying Wang is with the State Key Laboratory of Networking and Switching Technology, Beijing University of Posts and Telecommunications, Beijing 100876, China (wangying@bupt.edu.cn). Wen Chen is with the Department of Electronic Engineering, Shanghai Jiao Tong University, Shanghai 200240, China (e-mail: wenchen@sjtu.edu.cn).}\thanks{(Corresponding author: Zhou Su)}
\vspace{-1.5em}
}
\maketitle

\begin{abstract}
This paper investigates a reconfigurable intelligent surface (RIS)-assisted movable antenna (MA) secure integrated sensing and communication (ISAC) system. In this architecture, the RIS establishes indirect transmission links to provide communication services for multiple legitimate users, while the high spatial diversity gain of MA is leveraged to enhance system security. Then, we formulate an optimization problem to maximize the system total secrecy rate by jointly optimizing the MA position selection, active beamforming design for base station and passive beamforming design for RIS. The problem also accounts for practical constraints including transmit power budget, sensing beampattern mean square error (MSE), RIS unit-modulus constraint. However, it is challenging to solve this problem due to its non-convexity and strong coupling of the optimization variables. Consequently, we propose an alternating optimization (AO) framework, employing techniques including discrete binary particle swarm optimization (BPSO), successive convex approximation (SCA) and difference-of-convex (DC) programming to transform the optimization problem into convex subproblems. Based on the solution above, the convex sub-problems are solved iteratively until convergence is achieved. Numerical results demonstrate that the proposed algorithm outperforms other baseline algorithms in terms of secure communication performance.
\end{abstract}


\section{Introduction}
\IEEEPARstart{I}{n} recent years, accompanied by the continuous evolution of communication technologies, the sixth-generation (6G) mobile communication technology has gradually emerged as a core research focus in both academia and industry. As one of the key enabling technologies for 6G, integrated sensing and communication (ISAC) leverages its core advantages of hardware multiplexing, spectrum sharing, and functional synergy to support emerging application scenarios such as intelligent transportation and low-altitude economy\cite{ISAC1,ISAC2,ISAC3}. However, the practical deployment of ISAC still faces severe security challenge. 
In open physical channel environments, ISAC systems often face more eavesdropping threats than traditional communication systems due to the auto-correlation properties of ISAC waveforms and the coupling of sensing and communication information\cite{qieting1}\cite{qieting2}. Consequently, the security of ISAC systems has attracted more attention.

As an extension of ISAC, secure ISAC aims to mitigate potential communication threats through using the dual-functional nature of signals. The core objective of secure ISAC is to maximize the system's secrecy performance while strictly satisfying sensing constraints\cite{AN,xingzuo,learning,Pmax,UAV1,UAV2}. 
By jointly optimizing beamforming design and artificial noise (AN), \cite{AN} achieved significant gains in the system's secrecy performance. In \cite{learning}, the authors explored secure transmission problems in assisted computing scenarios based on machine learning algorithms. Considering the power consumption problem, \cite{Pmax} proposed a power control-based optimization algorithm to minimize energy overhead while ensuring secrecy performance. \cite{UAV1} extended the system to unmanned aerial vehicle (UAV) platforms, jointly optimized variables such as UAV trajectory and beamforming to maximize the total secrecy rate. However, it is worth noting that the above-mentioned studies primarily focus on scenarios where the direct link between the base station and the user is unobstructed. In the presence of the aforementioned blockage, the system faces the risk of significant degradation in secure communication performance.

To address the coverage blind spots caused by blockages, reconfigurable intelligent surface (RIS) has been widely adopted to reconfigure the wireless propagation environment\cite{RIS1.1}. By establishing virtual line-of-sight (LoS) links, RIS can effectively bypass obstacles, thereby restoring secure communication and serving capability in shadowed areas \cite{RIS1,RIS3,RIS4,satellite1,satellite2}. Existing research spans diverse deployment scenarios. In terrestrial networks, RIS is typically deployed on the surface of buildings to create indirect links. \cite{RIS1} proposed an active RIS-assisted ISAC architecture that maximizes the secrecy capacity under the sensing signal-to-interference-plus-noise (SINR) ratio constraint. The authors in \cite{RIS3} applied the majorization-minimization algorithm to solve the multi-user communication rate maximization problem in urban blockage environments. Extending to non-terrestrial networks, \cite{satellite1}\cite{satellite2} have embedded RIS-assisted secure ISAC systems into satellite scenarios. Specifically, \cite{satellite1} integrated RIS onto satellite platforms, leveraging high-orbit characteristics to build stable links and improve anti-eavesdropping robustness. \cite{satellite2} combined ISAC base stations with satellites, enhancing signal coverage through RIS reflection and jointly optimizing RIS phase shift and satellite orbits. However, the above works on RIS-assisted secure ISAC employ fixed position antennas (FPAs). Since an FPA relies on a static antenna layout, its spatial degrees of freedom (DoFs) are limited and cannot be adapted to favorable propagation conditions. Therefore, even with RIS assistance, the system’s potential in secrecy performance remains underutilized.

Movable antenna (MA) technology breaks the limitations of traditional FPA arrays, providing a novel technical pathway for secure ISAC systems.  Currently, hardware implementations of MA generally rely on either motor-driven mechanisms for continuous movement or fluid-based reconfigurable structures for discrete position switching\cite{MAtype1,MAtype2,MAtype3}. By flexibly adjusting antenna positions, the system equipped with MA is endowed with great beamforming DoFs and spatial diversity gains. Extensive research has been conducted to explore the potential of the MAs in wireless communication systems \cite{Li2,FRV,MAISAC1,MAISAC2,MAsecure1,MAsecure2,MAsecure3,MAbeam,MARIS1}.
\cite{FRV} established a field-response-based MA channel model and derived the theoretical performance limits of MA by constructing an MA-assisted wireless communication scenario. \cite{MAISAC1} and \cite{MAISAC2} introduced MA into ISAC systems. To be specific, \cite{MAISAC1} considered dynamic radar cross sections and utilized the flexibility of MA positions to reconstruct channels, validating the gain of MA on ISAC sensing performance, while \cite{MAISAC2} verified that MA-ISAC systems can significantly reduce the transmit power budget compared to FPA systems. Although the above-mentioned literature validates the important role of MA in enhancing sensing and communication performance, suppressing interference, and reducing transmit power from multiple dimensions, the secure communication performance of the MA-enabled ISAC system received limited attention. \cite{MAsecure1,MAsecure2,MAsecure3} further explored the research value of MA in secure ISAC systems. \cite{MAsecure1} provided a solution paradigm for secure communication problems in eavesdropping scenarios by jointly optimizing beamforming vectors and antenna positions. In \cite{MAsecure2}, the authors constructed a base station equipped with dual-panel MAs for full-duplex secure ISAC scenarios and proposed a multi-velocity particle swarm optimization algorithm to optimize MA positions. In scenarios facing complex environmental blockages, MA secure ISAC systems can establish indirect links through collaboration with RIS\cite{MAbeam,MARIS1}. In particular, to guarantee sensing fairness, \cite{MAbeam} achieved globally uniform sensing reliability by solving the max-min sensing beampattern optimization problem. The authors in \cite{MARIS1} introduced MA at the user end to deeply explore the secure transmission potential of MA in RIS-ISAC systems. 

Despite the demonstrated benefits of MAs in secure ISAC, existing studies mainly rely on the assumption of continuous antenna movement. In practical deployments, continuous MAs are typically driven by mechanical motors, which inherently introduce severe movement latency. Such delays may conflict with the short coherence time of wireless channels. Furthermore, continuous movement requires channel state information acquisition over a continuous spatial region, resulting in considerable estimation overhead. To bridge the gap between theoretical performance and hardware implementation, discrete MA architectures have emerged as a promising alternative. By enabling electronic switching among predefined candidate positions, discrete MA architectures can substantially reduce the CSI acquisition overhead and switching latency. However, this architectural shift fundamentally transforms the antenna-position design from a continuous optimization problem into a non-convex integer programming problem with discrete position variables, for which conventional continuous gradient-based methods cannot be directly applied \cite{r1c1}. Consequently, research on implementation-friendly discrete MA-enabled secure ISAC systems is crucial but remains in its infancy.

Based on the above discussion, to fully leverage the advantages of RIS and MA in enhancing signal coverage, improving channel DoFs, and spatial diversity gains, this paper constructs a RIS-assisted MA secure ISAC system aimed at maximizing the system total secrecy rate while satisfying sensing beampattern mean square error (MSE) constraint. Specifically, we consider a base station equipped with MA featuring discrete position selection, which injects AN to suppress the eavesdropper and provides secure communication services to multiple users with the assistance of RIS. Due to the high coupling among the optimization variables in the optimization problem, we propose a joint optimization algorithm to solve the problem of maximizing the total secrecy rate. The main contributions of this paper are as follows:

\begin{itemize}
\item This paper proposes a RIS-assisted MA secure ISAC system. Specifically, the RIS is employed to establish the communication link, while the spatial DoFs of the MA are exploited to enhance the system's secrecy performance. Then, a joint optimization problem is formulated with the MA position selection, beamforming design, AN, and RIS phase shift as optimization variables. This problem aims to maximize the total secrecy rate while satisfying the constraints on the base station transmit power budget, MA position selection, RIS phase shift, and sensing beampattern MSE. Given the strong coupling of the optimization variables and non-convexity of the optimization problem, direct solution poses a challenge.

\item To solve this non-convex optimization problem, we propose a joint discrete antenna positioning and beamforming design optimization algorithm. Specifically, based on the alternating optimization (AO) framework, we first determine the MA discrete position selection by applying the discrete binary particle swarm optimization (BPSO) algorithm. Subsequently, utilizing successive convex approximation (SCA) and difference of convex (DC) programming techniques, we handle the non-convex characteristics in the objective function and constraints to solve the active beamforming design for MA-BS and the passive beamforming for RIS, respectively. These three sub-problems are alternately optimized until convergence is achieved.

\item Numerical results demonstrate that the proposed discrete antenna positioning and beamforming design algorithm offers significant secrecy rate improvements compared to other baseline algorithms. Specifically, the RIS plays a crucial role in mitigating coverage blind spots, extending the system's reach and reshaping the propagation environment via precise passive beamforming. Moreover, by leveraging the high spatial DoFs from MA and AN injected by the base station, the system facilitates channel reconfiguration and interference suppression. 
\end{itemize}

The organization of this paper is as follows: Section \ref{sec2} introduces the system model of the RIS-assisted MA secure ISAC system and the formulation of the total secrecy rate maximization problem. The proposed joint discrete antenna positioning and beamforming design algorithm is elaborated in section \ref{sec3}. Section \ref{sec4} verifies the superiority of the proposed scheme through numerical results. Finally, Section \ref{sec5} concludes the full paper.

\textit{Notations:}  Scalars, vectors and matrices are denoted by lower-case letters, bold lower-case letters and bold upper-case letters, respectively. $\left| a \right|$ denotes the modulus of the scalar $a$, $\left\| {\bf{a}} \right\|$ and ${\left[ {\bf{a}} \right]_i}$ represent the norm and the $i$-th element of the vector ${\bf{a}}$, respectively. ${\rm{diag}}\left( {\bf{a}} \right)$ denotes a square matrix with the elements of ${\bf{a}}$ on its main diagonal. ${{\bf{A}}^*}$, ${{\bf{A}}^T}$, and ${{\bf{A}}^H}$ denote the conjugate, transpose, and conjugate transpose of the matrix ${\bf{A}}$, respectively. ${\bf{I}}$ denotes the identity matrix. ${\rm{rank}}\left( {\bf{A}} \right)$ and ${\rm{Tr}}\left( {\bf{B}} \right)$  represent the rank of matrix ${\bf{A}}$ and the trace of the square matrix ${\bf{B}}$, respectively. The $\left( {i,j} \right)$-th entry of matrix ${\bf{A}}$ is denoted by ${\left[ {\bf{A}} \right]_{i,j}}$. $ \otimes $ and $ \odot $ denote the Kronecker product and Hadamard product operators, respectively. $\left\langle {{\bf{X}},{\bf{Y}}} \right\rangle $ represents the Frobenius inner product of matrices ${\bf{X}}$ and ${\bf{Y}}$. ${\left\|  {\bf{A}}  \right\|_2}$ denotes the spectral norm (i.e., the maximum singular value) of the matrix ${\bf{A}}$. $\partial {\left\| {\bf{A}} \right\|_2}$ is the subgradient of matrix ${\bf{A}}$, and ${\bf{A}}\succeq0$ indicates that ${\bf{A}}$ is a positive semidefinite matrix. ${\mathbb{C}^{a \times b}}$ denotes the set of $a \times b$ complex matrices. ${\cal C}{\cal N}\left( {0,b} \right)$ represents the circularly symmetric complex Gaussian distribution with zero mean and covariance $b$. ${\mathbb{E}}[\cdot]$ denotes the expectation operation, and $[\cdot]^+$represents $\max(\cdot,0)$ operation.

\section{System Model and Problem Formulation}\label{sec2}
\begin{figure}[t]
    \centering
    \includegraphics[height=55mm, keepaspectratio]{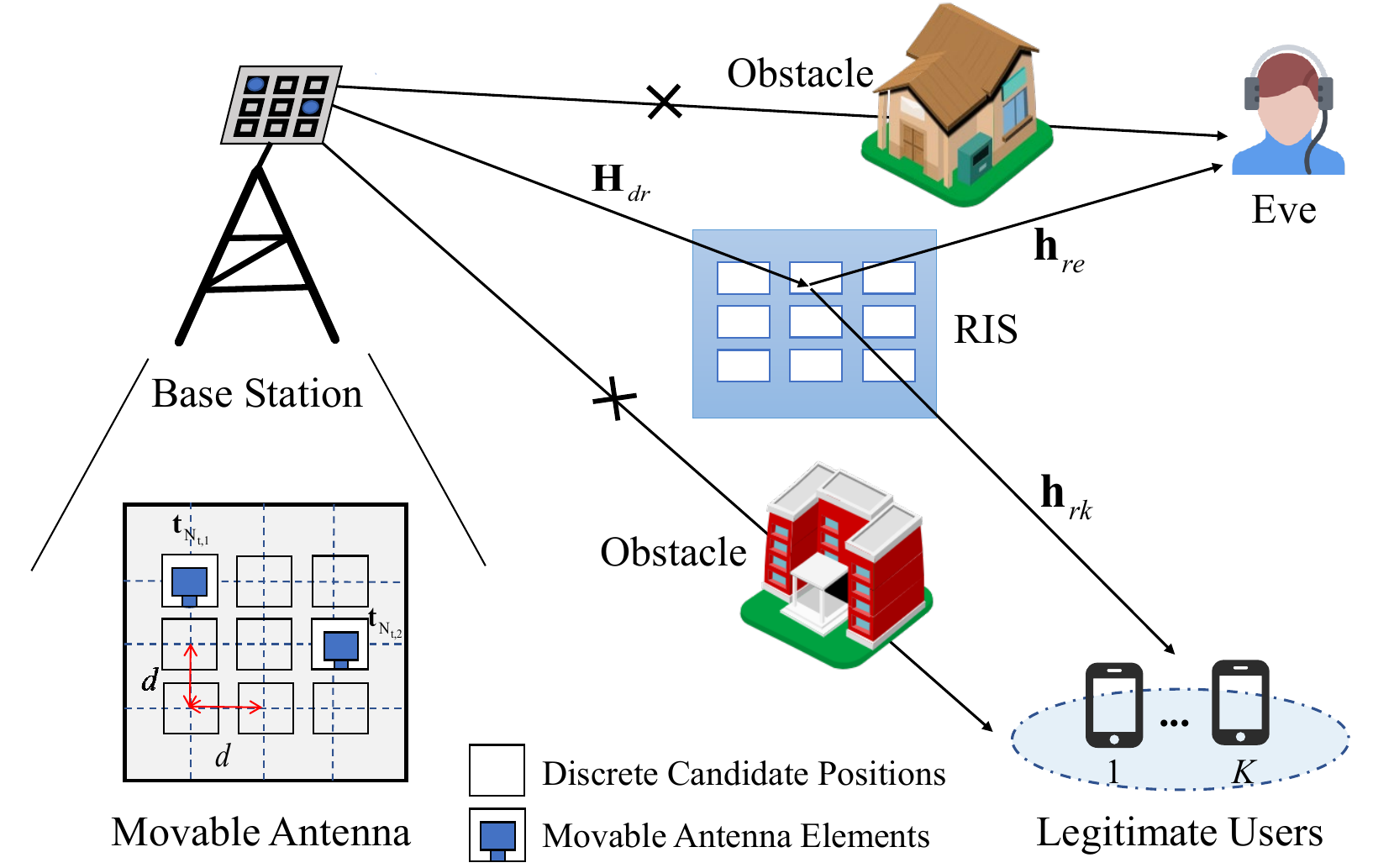}
    \caption{Illustration of RIS-assisted MA secure ISAC systems.}
    \label{fig1}
    \vspace{2em}
\end{figure}

This section presents the system model for a RIS-assisted MA secure ISAC system, as illustrated in Fig. \ref{fig1}. The system comprises a dual-functional radar and communication base station equipped with MA (hereinafter referred to as the MA-BS), a RIS, $K$ single-antenna legitimate users and a single-antenna potential eavesdropper. The MA-BS is equipped with a panel containing ${N_{\rm{t}}}$ MA elements, which features $M$ discrete candidate positions. The RIS is deployed on the surface of a building and consists of $T = {T_x} \times {T_y}$ reflecting elements. It is assumed that both the discrete candidate of the MA and the reflecting elements of the RIS are arranged in a uniform planar array (UPA) geometry. We consider a scenario where the direct links are completely blocked. Consequently, the MA-BS relies on the assistance of the RIS to establish communication links with legitimate users, while simultaneously exploiting the signals reflected by the RIS to further enhance system security.

The $M$ discrete candidate positions are collected into the position matrix ${\bf{P}} = \left[ {{{\bf{p}}_1}, \cdots {{\bf{p}}_M}} \right]$, where ${{\bf{p}}_m} = \left[ {{x_m},{y_m}} \right]$ denotes the $m$-th candidate position. Let ${{\bf{b}}_{{n_{\rm{t}}}}} = {\left[ {{b_{{n_{\rm{t}}}}}\left[ 1 \right], \cdots {b_{{n_{\rm{t}}}}}\left[ M \right]} \right]^T}$ be the binary position selection vector for the ${n_{\rm{t}}}$-th MA elements. Note that each MA element can select a specific discrete candidate position, for $\forall m \in \left\{ {1, \cdots ,M} \right\}$, the constraints ${b_{{n_{\rm{t}}}}}\left[ m \right] \in \left\{ {0,1} \right\}$ and $\sum\nolimits_{m = 1}^M {{b_{{n_{\rm{t}}}}}\left[ m \right]}  = 1,\forall {n_{\rm{t}}}$ must be satisfied. Specially, ${b_{{n_{\rm{t}}}}}\left[ m \right] = 1$ if and only if the $m$-th candidate position is occupied by the ${n_{\rm{t}}}$-th MA element. Consequently, the actual position of the ${n_{\rm{t}}}$-th MA element can be expressed as ${{\bf{t}}_{{n_{\rm{t}}}}} = {\left[ {{t_{{t_x},{n_{\rm{t}}}}},{t_{{t_y},{n_{\rm{t}}}}}} \right]^T}$. Furthermore, we define the distance matrix ${\bf{D}} \in \mathbb{C}^{M \times M}$, where the entry ${D_{m,m'}}$ represents the Euclidean distance between the $m$-th and $m'$-th candidate positions. Given that multiple MA elements cannot be deployed at the same candidate position, any pair of MA elements must satisfy the following minimum distance constraint
\begin{equation}
{\bf{b}}_{{n_{\rm{t}}}}^T{\bf{D}}{{\bf{b}}_{{n_{\rm{t}}}'}} \ge {D_{\min }},{n_{\rm{t}}} \ne {n_{\rm{t}}}',\forall {n_{\rm{t}}},{n_{\rm{t}}}',
\end{equation}
where ${D_{\min }}$ denotes the minimum allowable distance between any pair of MA elements.

In this system, the RIS is utilized to establish communication links and provide signal coverage when the LoS paths between the BS and users are blocked. The reflection coefficient matrix of the RIS is expressed as
\begin{equation}
    {\bf{\Theta }} = {\rm{diag}}\left\{ {{e^{j{\alpha _1}}}, \ldots ,{e^{j{\alpha _t}}}, \ldots ,{e^{j{\alpha _T}}}} \right\},
\end{equation}
where ${\alpha _t} \in \left[ {0,2\pi } \right)$, $\forall t$ represents the phase shift of the $t$-th reflecting element. Due to severe path loss, signal scattering involving two or more hops at the RIS is negligible\cite{Li3}. To maximize the strength of the reflected signals, the RIS phase shift matrix must satisfy the following unit-modulus constraint
\begin{equation}
    \left| {{{\left[ {\bf{\Theta }} \right]}_{t,t}}} \right| = {\rm{1,}}\forall t.
\end{equation}
        
\subsection{Channel Model}
We assume that all wireless channels in the RIS-assisted MA secure ISAC system are planar far-field channels, where each transmit and receive path within the same cluster shares identical angles of departure (AoDs), angles of arrival (AoAs), and path fading coefficients. Since the direct links between the MA-BS and the users are completely blocked, the MA-BS needs the assistance of the RIS to provide secure communication services to legitimate users and to perform sensing on the potential eavesdropper. Considering the multipath scattering effects in practical propagation environments, the dominant path component remains significant even in the presence of obstructions between the MA-BS and the receivers. Consequently, all channels are modeled as Rician fading channels. Given the UPA structure of the RIS, its array response vector can be expressed as
\begin{equation}
    {{\bf{a}}_{{\rm{horiz}}}}\left( {{\theta _{\rm{s}}},{\phi _{\rm{s}}}} \right) = {\left[ {1, \cdot  \cdot  \cdot ,{e^{j\left( {{T_x} - 1} \right)\frac{{2\pi }}{\lambda }{d_{\rm{1}}}\sin \left( {{\phi _{\rm{s}}}} \right)\cos \left( {{\theta _{\rm{s}}}} \right)}}} \right]^T} \!\in\! {\mathbb{C}^{{T_x} \times 1}},
\end{equation}
\begin{equation}
    {{\bf{a}}_{{\rm{vert}}}}\left( {{\theta _{\rm{s}}}} \right) = {\left[ {1, \cdot  \cdot  \cdot ,{e^{j\left( {{T_y} - 1} \right)\frac{{2\pi }}{\lambda }{d_{\rm{2}}}\sin \left( {{\theta _{\rm{s}}}} \right)}}} \right]^T} \in {\mathbb{C}^{{T_y} \times 1}},
\end{equation}
where ${d_1}$ and ${d_2}$ denote the horizontal and vertical spacing between adjacent RIS elements, respectively, which are typically set to ${\lambda  \mathord{\left/
 {\vphantom {\lambda  2}} \right.
 \kern-\nulldelimiterspace} 2}$. Additionally, ${\theta _{\rm{s}}}$ and ${\phi _{\rm{s}}}$ represent the vertical and horizontal angles from the RIS to the users (including both legitimate users and the eavesdropper), respectively. Accordingly, the LoS components for the channels from the RIS to the $k$-th legitimate user and from the RIS to the eavesdropper are given by
 \begin{equation}
    {\bf{h}}_{rk}^{{\rm{LoS}}} = {{\bf{a}}_{{\rm{horiz}}}}\left( {{\theta _k},{\phi _k}} \right) \otimes {{\bf{a}}_{{\rm{vert}}}}\left( {{\theta _k}} \right) \in {\mathbb{C}^{T \times 1}},
 \end{equation}
 \begin{equation}
      {\bf{h}}_{re}^{{\rm{LoS}}} = {{\bf{a}}_{{\rm{horiz}}}}\left( {{\theta _e},{\phi _e}} \right) \otimes {{\bf{a}}_{{\rm{vert}}}}\left( {{\theta _e}} \right) \in {\mathbb{C}^{T \times 1}}.
 \end{equation}
 Thus, the channel between the RIS and the $k$-th legitimate user can be expressed as
 \begin{equation}
     {{\bf{\bar h}}_{rk}} = \sqrt {\frac{\vartheta }{{1 + \vartheta }}} {\bf{h}}_{rk}^{{\rm{LoS}}} + \sqrt {\frac{1}{{1 + \vartheta }}} {\bf{h}}_{rk}^{{\rm{NLoS}}},\forall k,
 \end{equation}
 where ${\left[{{\bf{h}}_{rk}^{{\rm{NLoS}}}}\right]_j}\sim{\cal C}{\cal N}\left( {0,1} \right)$, and $\vartheta $ denotes the Rician factor of this channel. Similarly, the channel from the RIS to the eavesdropper is formulated as
 \begin{equation}
     {\overline {\bf{h}} _{re}} = \sqrt {\frac{\eta }{{1 + \eta }}} {\bf{h}}_{re}^{{\rm{LoS}}} + \sqrt {\frac{1}{{1 + \eta }}} {\bf{h}}_{re}^{{\rm{NLoS}}},
 \end{equation}
 where ${\left[ {{\bf{h}}_{re}^{{\rm{NLoS}}}} \right]_j}\sim{\cal C}{\cal N}\left( {0,1} \right)$, and $\eta $ represents the corresponding Rician factor. The corresponding channel gains are modeled as follows
 \begin{equation}
     {{\bf{h}}_{rk}} = \sqrt {{C_0}{{\left( {\frac{{{d_{rk}}}}{{{D_0}}}} \right)}^{ - \varpi }}} {{\bf{\bar h}}_{rk}},
 \end{equation}
\begin{equation}
    {{\bf{h}}_{re}} = \sqrt {{C_0}{{\left( {\frac{{{d_{re}}}}{{{D_0}}}} \right)}^{ - \iota }}} {{\bf{\bar h}}_{re}},
\end{equation}
where ${C_0}$ denotes the path loss at the reference distance ${D_0} = 1$ m, and $\varpi $ and $\iota $ represent the path loss exponents of the respective channels. Additionally, ${d_{rk}}$ and ${d_{re}}$ denote the distance between the RIS and the $k$-th legitimate user and that between the RIS and the eavesdropper, respectively.

The channel between the MA-BS and the RIS depends on the position selection of the MA. Its channel gain matrix is denoted as ${{\bf{ H}}_{dr}} = {\widehat {\bf{H}}_{dr}}{{\bf{B}}_{\rm{t}}} \in {\mathbb{C}^{T \times {N_{\rm{t}}}}}$, where ${\widehat {\bf{H}}_{dr}} = \left[ {{{\widehat {\bf{H}}}_1}, \cdots ,{{\widehat {\bf{H}}}_{{N_{\rm{t}}}}}} \right] \in {\mathbb{C}^{T \times M{N_{\rm{t}}}}}$ represents the channel gain matrix between the $T$ RIS reflecting elements and the MA. Specifically, ${\widehat {\bf{H}}_{{n_{\rm{t}}}}} = \left[ {{{\bf{h}}_{{n_{\rm{t}}}}}\left( {{{\bf{p}}_1}} \right), \cdots ,{{\bf{h}}_{{n_{\rm{t}}}}}\left( {{{\bf{p}}_M}} \right)} \right] \in {\mathbb{C}^{T \times M}}$ denotes the channel vector between the $T$ RIS reflecting elements and the ${n_{\rm{t}}}$-th MA element. ${{\bf{h}}_{{n_{\rm{t}}}}}\left( {{{\bf{p}}_m}} \right){\rm{ = }}{\left[ {{h_{{n_{\rm{t}}},1}}\left( {{{\bf{p}}_m}} \right), \cdots ,{h_{{n_{\rm{t}}},T}}\left( {{{\bf{p}}_m}} \right)} \right]^T} \in {\mathbb{C}^{T \times 1}}$ represents the channel between the ${n_{\rm{t}}}$-th MA element located at ${{\bf{p}}_m}$ and the RIS. The channel coefficient between the ${n_{\rm{t}}}$-th MA element at position ${{\bf{p}}_m}$ and the $t$-th RIS reflecting element, denoted as ${h_{{n_{\rm{t}}},t}}\left( {{{\bf{p}}_m}} \right) \in \mathbb{C}$, is formulated as
\vspace{1em}
\begin{equation}
    {h_{{n_{\rm{t}}},t}}\left( {{{\bf{p}}_m}} \right) = \sqrt \alpha  {e^{\frac{{j2\pi \left( {\left( {{x_m} - {x_1}} \right)\cos \theta _{h,t}^T\sin \phi _{h,t}^T + \left( {{y_m} - {y_1}} \right)\sin \theta _{h,t}^T} \right)}}{\lambda }}},
\end{equation}
where $\alpha $ denotes the channel fading coefficient between the MA-BS and the RIS, $\theta _{h,t}^T$ and $\phi _{h,t}^T$ represent the elevation and azimuth angles of the channel path at the $t$-th RIS reflecting element, respectively. The position selection matrix of the MA ${{\bf{B}}_{\rm{t}}} \in {\mathbb{C}^{{N_{\rm{t}}}M \times {N_{\rm{t}}}}}$ is defined as
\begin{equation}
    {{\bf{B}}_{\rm{t}}} = \left[ {\begin{array}{*{20}{c}}
{{{\bf{b}}_1}}&{{{\bf{0}}_M}}& \cdots &{{{\bf{0}}_M}}\\
{{{\bf{0}}_M}}&{{{\bf{b}}_2}}& \cdots &{{{\bf{0}}_M}}\\
 \vdots & \vdots & \ddots & \vdots \\
{{{\bf{0}}_M}}&{{{\bf{0}}_M}}& \cdots &{{{\bf{b}}_{{N_{\rm{t}}}}}}
\end{array}} \right].
\end{equation}

\subsection{Signal Model}
In this paper, the MA-BS transmits a dual-functional sensing and communication signal through multi-antenna beamforming, which is expressed as
\begin{equation}
    {\bf{x}} = \sum\limits_{k = 1}^K {{{\bf{w}}_k}} {s_k} + {\bf{z}},
\end{equation}
where ${{\bf{w}}_k} \in {\mathbb{C}^{{N_{\rm{t}}} \times 1}}$ denotes the beamforming vector for the $k$-th legitimate user, and ${s_k}$ represents the information-bearing symbol intended for the $k$-th legitimate user. To further enhance security, an AN signal ${\bf{z}} \in {\mathbb{C}^{{N_{\rm{t}}} \times 1}}$ is introduced, satisfying ${\left[ {\bf{z}} \right]_i} \sim {\cal C}{\cal N}\left( {0,1} \right),\forall i$. Its covariance matrix can be expressed as ${\bf{R}} = \mathbb{E}\left[ {{\bf{z}}{{\bf{z}}^H}} \right]$. The useful signal and the AN signal are statistically independent. Consequently, the covariance matrix of the transmitted signal can be formulated as ${{\bf{R}}_x} = \mathbb{E}\left[ {{\bf{x}}{{\bf{x}}^H}} \right] = \sum\nolimits_{k = 1}^K {{{\bf{w}}_k}{\bf{w}}_k^H + {\bf{R}}}$. Therefore, the received signal at the $k$-th legitimate user is given by

\begin{equation}
    y_k=\sum_{c=1}^{K}
    \mathbf{h}_{rk}^{H}\mathbf{\Theta}\mathbf{H}_{dr}
    \mathbf{w}_{c}s_{c}
    +\mathbf{h}_{rk}^{H}\mathbf{\Theta}\mathbf{H}_{dr}\mathbf{z}
    +n_k, \forall k,
\end{equation}
where ${n_k} \sim {\cal C}{\cal N}\left( {0,\sigma _k^2} \right)$ represents the additive white Gaussian noise (AWGN) introduced at the
receiving antenna of the $k$-th legitimate user. The signal received at the eavesdropper is expressed as 
\begin{equation}
    {y_e} = \sum\limits_{k = 1}^K {{\bf{h}}_{re}^H{\bf{\Theta }}{{\bf{H}}_{dr}}{{\bf{w}}_k}{s_k}}  + {\bf{h}}_{re}^H{\bf{\Theta }}{{\bf{H}}_{dr}}{\bf{z}} + {n_e},
\end{equation}
where ${n_e} \sim {\cal C}{\cal N}\left( {0,\sigma _e^2} \right)$ denotes the AWGN introduced at the eavesdropper.
To ensure high-quality sensing of the eavesdropper, it is necessary to illuminate the desired target location with a beam that exhibits strong energy focusing and low sidelobe leakage, thereby effectively distinguishing the desired echo from clutter. To this end, the elevation domain $[-\pi/2, \pi/2]$ and the azimuth domain $[-\pi/2, \pi/2]$ are discretized into $L$ and $Q$ directions, respectively. The ideal beampattern can be expressed as
\begin{equation}
    {\cal D}\!\left( {{\theta _l},{\phi _q}} \right) = 
    \begin{cases}
        1, & \begin{aligned}[t]
                &\theta_e - \Delta \le \theta_l \le \theta_e + \Delta \\
                &\text{and } \phi_e - \delta \le \phi_q \le \phi_e + \delta
              \end{aligned}\\
        0, & \text{otherwise}
    \end{cases},
\end{equation}
where $2\Delta $ and $2\delta$ represent the beamwidths of the target in the elevation and azimuth directions, ${\theta _l}$ and ${\phi _q}$ are the elevation and azimuth angles of the eavesdropper from the perspective of the RIS. To quantify the beampattern matching accuracy, the MSE between the ideal beampattern and the actual beampattern is adopted as the sensing performance metric, which is formulated as
\begin{align}
  {\cal M}\left( {{\theta _l},{\phi _q}} \right)=&\frac{1}{QL}\sum\limits_{q=1}^Q \sum\limits_{l=1}^L\bigg| {\rho_0}{\cal D}\left({{\theta_l},{\phi_q}}\right) - {{\widehat{\bf{a}}}^H}\left( {{\theta _l},{\phi _q}}\right) \nonumber \\
        & \times {\bf{\Theta }}{{\bf{H}}_{dr}}{{\bf{R}}_x}{\bf{H}}_{dr}^H{{\bf{\Theta }}^H}\widehat {\bf{a}}\left( {{\theta _l},{\phi _q}} \right) \bigg|^2 ,
\end{align}
where ${\rho _0}$ is a scaling factor, and $\widehat {\bf{a}}\left( {{\theta _l},{\phi _q}} \right)$ denotes the field response vector at the RIS, specifically expressed as
\vspace{1em}
\begin{equation}
    \widehat {\bf{a}}\left( {{\theta _l},{\phi _q}} \right) = {{\bf{a}}_{{\rm{horiz}}}}\left( {{\theta _l},{\phi _q}} \right) \otimes {{\bf{a}}_{{\rm{vert}}}}\left( {{\theta _l}} \right) \in {\mathbb{C}^{T \times 1}}.
\end{equation}

\subsection{Problem Formulation}
The secrecy rate is one of the crucial metrics for evaluating the secure communication performance of communication systems. The secrecy rate between the MA-BS and legitimate users is determined by both their achievable communication rate and the eavesdropping rate of the eavesdropper on their communication. According to the Shannon formula, the communication rate between the MA-BS and the $k$-th legitimate user can be expressed as
\begin{equation}
    {R_k} = {\log _2}\left( {1 + {\gamma _k}} \right),\forall k,
\end{equation}
where ${\gamma _k}$ denotes the SINR at the $k$-th legitimate user, which is specially formulated as
\begin{equation}
    {\gamma _k} = \frac{{{{\left| {{\bf{h}}_{rk}^H{\bf{\Theta }}{{\bf{H}}_{dr}}{{\bf{w}}_k}} \right|}^2}}}{{\sum\limits_{c \ne k}^K {{{\left| {{\bf{h}}_{rk}^H{\bf{\Theta }}{{\bf{H}}_{dr}}{{\bf{w}}_c}} \right|}^2} \!+\! {\bf{h}}_{rk}^H{\bf{\Theta }}{{\bf{H}}_{dr}}{\bf{RH}}_{dr}^H{{\bf{\Theta }}^H}{{\bf{h}}_{rk}} \!+\! \sigma _k^2} }}.
\end{equation}

Assuming that the eavesdropper can eliminate interference from other users before decoding the information of a specific legitimate user, the eavesdropping rate when the eavesdropper intercepts the communication between the MA-BS and the $k$-th legitimate user can be written as
\vspace{0.5em}
\begin{equation}
    {C_k} = {\log _2}\left( {1 + \frac{{{{\left| {{\bf{h}}_{re}^H{\bf{\Theta }}{{\bf{H}}_{dr}}{{\bf{w}}_k}} \right|}^2}}}{{{\bf{h}}_{re}^H{\bf{\Theta }}{{\bf{H}}_{dr}}{\bf{RH}}_{dr}^H{{\bf{\Theta }}^H}{{\bf{h}}_{re}} + \sigma _e^2}}} \right),\forall k.
\end{equation}
Therefore, the achievable secrecy rate between the MA-BS and the $k$-th legitimate user is given by
\vspace{0.5em}
\begin{equation}
    {D_k} = {\left[ {{R_k} - {C_k}} \right]^ + },\forall k.
\end{equation}

The single eavesdropper model with known channel information is adopted as a benchmark for evaluating the secrecy performance of the considered system.\footnote{Since the BS and legitimate users have an authorized association, a detected but unauthorized terminal can be regarded as a potential eavesdropper. Its angular or positional information may be estimated through ISAC beam scanning, providing a basis for approximating the corresponding channel. Extensions to multiple eavesdroppers and imperfect eavesdropper CSI will be considered in future work.}
To achieve secure communication and effective sensing, this paper aims to maximize the total secrecy rate while satisfying the constraints on transmit power budget, MA position selection, MA minimum distance, RIS phase shift and sensing beampattern MSE. The specific optimization problem (P0) can be formulated as follows
\allowdisplaybreaks
\begin{subequations}
    \label{P0}
    \renewcommand{\theequation}{24\alph{equation}}
    \begin{align}
        \text{(P0):}\mathop {\max }\limits_{\left\{ {{{\bf{w}}_k}} \right\},{\bf{z}},{\bf{\Theta }},{{\bf{B}}_{\rm{t}}}} & \quad  \sum\limits_{k = 1}^K {{D_k}},\notag \\
        \text{s.t.}\quad
        & \sum\limits_{k = 1}^K {{{\left\| {{{\bf{w}}_k}} \right\|}^2}}  + {\left\| {\bf{z}} \right\|^2} \le {P_{\max }},\label{P}\\
        & \sum\limits_{m = 1}^M {{b_{{n_{\rm{t}}}}}\left[ m \right]}  = 1,\label{MA1}\\
        & {b_{{n_{\rm{t}}}}}\left[ m \right] \in \left\{ {0,1} \right\},\forall {n_{\rm{t}}},m,\label{MA2} \\
        & {\bf{b}}_{{n_{\rm{t}}}}^T{\bf{D}}{{\bf{b}}_{{n_{\rm{t}}}'}}\! \ge\! {D_{\min }},{n_{\rm{t}}} \!\ne\! {n_{\rm{t}}}',\forall {n_{\rm{t}}},{n_{\rm{t}}}',\label{MA3}\\
        & \left| {{{\left[ {\bf{\Theta }} \right]}_{t,t}}} \right| = {\rm{1,}}\forall t,\label{RIS}\\
        & 
            \frac{1}{QL}\sum\limits_{q = 1}^Q \sum\limits_{l = 1}^L \bigg| {\rho _0}{\cal D}\left( {{\theta _l},{\phi _q}} \right) - {{\widehat {\bf{a}}}^H}\left( {{\theta _l},{\phi _q}} \right) \nonumber \\
            & \times {\bf{\Theta }}{{\bf{H}}_{dr}}{{\bf{R}}_x}{\bf{H}}_{dr}^H{{\bf{\Theta }}^H}\widehat {\bf{a}}\left( {{\theta _l},{\phi _q}} \right) \bigg|^2 \!\le\! \varepsilon.\label{sensing}
    \end{align}
\end{subequations}
Specifically, constraint \eqref{P} represents the transmit power constraint, where ${P_{\max }}$ denotes the maximum transmit power of the MA-BS. Constraints \eqref{MA1} and \eqref{MA2} ensure that the MA satisfies the discrete position selection requirements, while \eqref{MA3} enforces the minimum spacing constraint between MAs. Constraint \eqref{RIS} corresponds to the reflection coefficient constraint of the RIS. \eqref{sensing} indicates the sensing constraint, where $\varepsilon$ represents the sensing threshold. The sensing constraint introduces a tradeoff between secure communication and sensing performance. A smaller $\varepsilon$ imposes a stricter sensing requirement and limits secrecy rate optimization, while a larger threshold relaxes the sensing constraint and provides more flexibility for secure transmission.


\section{Joint Discrete Antenna Positioning and Beamforming Design Optimization Algorithm}\label{sec3}
It is challenging to directly solve the problem (P0) due to its non-convex nature, which includes the fractional structure of the objective function and the non-convex constraints \eqref{MA2}-\eqref{sensing}, as well as the coupling among the optimization variables $\{{\bf{w}}_k\},\forall{k}$, ${\bf{z}}$, ${\bf{\Theta }}$ and ${{\bf{B}}_{\rm{t}}}$. To facilitate the solution, we first transform the objective function of (P0) using a logarithmic operation, which can be expressed as
\begin{equation}
    \sum\limits_{k = 1}^K \begin{array}{l}
{\log _2}\left( {\sum\limits_{c = 1}^K {{{\left| {{{\bf{h}}_k}{{\bf{w}}_c}} \right|}^2}}  + {{\left| {{{\bf{h}}_k}{\bf{z}}} \right|}^2} + \sigma _k^2} \right) - \\{\log _2}\left( {\sum\limits_{c \ne k}^K {{{\left| {{{\bf{h}}_k}{{\bf{w}}_c}} \right|}^2}}  + {{\left| {{{\bf{h}}_k}{\bf{z}}} \right|}^2} + \sigma _k^2} \right) - \\
{\log _2}\left( {{{\left| {{{\bf{h}}_e}{{\bf{w}}_k}} \right|}^2} \!+\! {{\left| {{{\bf{h}}_e}{\bf{z}}} \right|}^2} \!+\! \sigma _e^2} \right) \!+\! {\log _2}\left( {{{\left| {{{\bf{h}}_e}{\bf{z}}} \right|}^2} \!+\! \sigma _e^2} \right),
\end{array} 
\end{equation}
where ${{\bf{h}}_k}\!=\!{\bf{h}}_{rk}^H{\bf{\Theta }}{{\bf{H}}_{dr}} \!\in \!{\mathbb{C}^{1 \times {N_{\rm{t}}}}},\forall k$, ${{\bf{h}}_e} = {\bf{h}}_{re}^H{\bf{\Theta }}{{\bf{H}}_{dr}} \in {\mathbb{C}^{1 \times {N_{\rm{t}}}}}$. By applying the matrix lifting technique, let ${{\bf{H}}_k} = {\bf{h}}_k^H{{\bf{h}}_k} \in {\mathbb{C}^{{N_{\rm{t}}} \times {N_{\rm{t}}}}},\forall k$, ${{\bf{H}}_e} = {\bf{h}}_e^H{{\bf{h}}_e} \in {\mathbb{C}^{{N_{\rm{t}}} \times {N_{\rm{t}}}}}$, ${{\bf{W}}_k} = {{\bf{w}}_k}{\bf{w}}_k^H \in {\mathbb{C}^{{N_{\rm{t}}} \times {N_{\rm{t}}}}},\forall k$ and ${\bf{Z}} = {\bf{z}}{{\bf{z}}^H} \in {\mathbb{C}^{{N_{\rm{t}}} \times {N_{\rm{t}}}}}$. Based on matrix trace operations, the objective function can be further reformulated as \eqref{F1}.
\begin{figure*}[t]
\begin{equation}
\mathcal{F}(\{\mathbf{W}_k\}, \mathbf{Z}, \boldsymbol{\Theta}, \mathbf{B}_t) = \sum\limits_{k = 1}^K \left(
\begin{aligned}
& \underbrace{\log_2 \left( \sum_{c=1}^{K} \text{Tr}(\mathbf{H}_k \mathbf{W}_c) + \text{Tr}(\mathbf{H}_k \mathbf{Z}) + \sigma_k^2 \right)}_{R_k^1} - \\
& \underbrace{\log_2 \left( \sum_{ c \neq k}^{K} \text{Tr}(\mathbf{H}_k \mathbf{W}_c) + \text{Tr}(\mathbf{H}_k \mathbf{Z}) + \sigma_k^2 \right)}_{R_k^2} - \\
& \underbrace{\log_2 \left( \text{Tr}(\mathbf{H}_e \mathbf{W}_k) + \text{Tr}(\mathbf{H}_e \mathbf{Z}) + \sigma_e^2 \right)}_{C_k^1} + \underbrace{\log_2 \left( \text{Tr}(\mathbf{H}_e \mathbf{Z}) + \sigma_e^2 \right)}_{C_k^2}
\end{aligned}
\right)
\label{F1}
\end{equation}
\hrule
\end{figure*}

Consequently, the optimization problem (P0) can be modified into the following problem (P1):

\allowdisplaybreaks
\begin{subequations}
\label{P1}
\renewcommand{\theequation}{27\alph{equation}} 
\begin{align}
\text{(P1):}\mathop {\max }\limits_{\{{{\bf{W}}_k}\},{\bf{Z}},{\bf{\Theta }},{{\bf{B}}_{\rm{t}}}} &{\cal F}\left( {\{{{\bf{W}}_k}\}{\rm{,}}{\bf{Z}}{\rm{,}}{\bf{\Theta }}{\rm{,}}{{\bf{B}}_{\rm{t}}}} \right),\nonumber\\
\text{s.t.}\quad
& \sum_{k = 1}^K {\rm{Tr}}\left( \mathbf{W}_k \right) + {\rm{Tr}}\left( \mathbf{Z} \right) \le P_{\max}, \\
& \sum_{m = 1}^M b_{n_{\rm{t}}}[m] = 1, \\
& b_{n_{\rm{t}}}[m] \in \{0,1\},\forall n_{\rm{t}},m, \\
& \mathbf{b}_{n_{\rm{t}}}^T{\bf{D}}\mathbf{b}_{n_{\rm{t}}'} \ge D_{\min},n_{\rm{t}} \ne n_{\rm{t}}',\forall n_{\rm{t}},n_{\rm{t}}', \label{dist1} \\
& \left| \left[ \mathbf{\Theta} \right]_{t,t} \right| = 1,\forall t, \\
&\frac{1}{QL}\sum\limits_{q = 1}^Q \sum\limits_{l = 1}^L \bigg| {\rho _0}{\cal D}\left( {{\theta _l},{\phi _q}} \right) - {{\widehat {\bf{a}}}^H}\left( {{\theta _l},{\phi _q}} \right) \nonumber \\
& \times {\bf{\Theta }}{{\bf{H}}_{dr}}{{\bf{R}}_x}{\bf{H}}_{dr}^H{{\bf{\Theta }}^H}\widehat {\bf{a}}\left( {{\theta _l},{\phi _q}} \right) \bigg|^2 \!\le\! \varepsilon.\\
& \mathbf{W}_k \succeq 0, {\rm{rank}}\left( \mathbf{W}_k \right) = 1,\forall k, \label{WK}\\
& \mathbf{Z} \succeq 0, {\rm{rank}}\left( \mathbf{Z} \right) = 1.\label{Z}
\end{align}
\end{subequations}
The application of the matrix lifting technique introduces additional positive semidefinite and rank-one constraints (i.e., \eqref{WK} and \eqref{Z}) to the optimization problem. To address the non-convex problem, we propose a joint discrete antenna positioning and beamforming design algorithm based on the AO framework, as described in the following subsections. Specifically, the optimization problem (P1) is decomposed into three sub-problems. First, given the MA-BS beamforming design, AN and RIS phase shift, we propose a BPSO-based method for MA position optimization. Subsequently, by fixing the MA position selection and the RIS phase shift, we develop an active beamforming design algorithm for the MA-BS. Finally, the RIS phase shift matrix is optimized given the MA position selection, beamforming design and AN.

\subsection{BPSO-Based MA Position Optimization}
The BPSO algorithm utilizes the collaboration of discrete particles within a swarm to rapidly approach high-quality feasible solutions. Taking all factors into account, we apply the BPSO algorithm to solve the proposed problem (P1). First, the beamforming design $\left\{ {{{\bf{W}}_k}} \right\},\forall k$, the AN matrix ${\bf{Z}}$ and the RIS phase shift  ${\bf{\Theta }}$ are fixed. Following the BPSO framework, the positions of $I$ particles are initialized as ${{\cal G}^{(0)}} = \left\{ {{\bf{\bar g}}_1^{(0)},{\bf{\bar g}}_2^{(0)},...,{\bf{\bar g}}_I^{(0)}} \right\}$, where each particle ${\bf{\bar g}}_i^{\left( 0 \right)} = {\left[ {{\bf{g}}_1^T,{\bf{g}}_2^T, \cdots ,{\bf{g}}_{{N_{\rm{t}}}}^T} \right]^T} \in {\mathbb{C}^{M{N_{\rm{t}}} \times {\rm{1}}}}$ represents a possible distribution of the MA elements. The velocities of the $I$ particles are initialized as ${{\cal V}^{(0)}} = \left\{ {{\bf{v}}_1^{(0)},{\bf{v}}_2^{(0)}, \ldots ,{\bf{v}}_I^{(0)}} \right\}$, where ${\bf{v}}_i^{(0)} = {\left[ {{\bf{v}}_1^T,{\bf{v}}_2^T, \ldots ,{\bf{v}}_{{N_{\rm{t}}}}^T} \right]^T} \in {\mathbb{C}^{M{N_{\rm{t}}} \times 1}}$, ${\bf{v}}_{{n_{\rm{t}}}}^T \in {\mathbb{C}^{M \times 1}}$ denotes the velocity of the ${n_{\rm{t}}}$-th MA element across different candidate positions, satisfying ${v_{{\rm{min}}}} \le {\bf{v}}_{{n_{\rm{t}}}}^T\left( m \right) \le {v_{{\rm{max}}}}$. In the $j$-th iteration, the velocity of each particle is updated as
\begin{align}
        {\bf{v}}_i^{(j)}\left( d \right) = &\xi {\bf{v}}_i^{(j - 1)}\left( d \right) + {c_1}{r_1}\left( {{\bf{\bar g}}_i^ * \left( d \right) - {\bf{\bar g}}_i^{(j - 1)}\left( d \right)} \right) \nonumber \\ 
        &+ {c_2}{r_2}\left( {{{{\bf{\bar g}}}^ * }\left( d \right) - {\bf{\bar g}}_i^{(j - 1)}\left( d \right)} \right),
\end{align}
where the iteration index $j$ satisfies $0 \le j \le J$, and $\xi$ is the inertia weight, expressed as 
$\xi  = {\xi _{\max }} - j({\xi _{\max }} - {\xi _{\min }})/J$ typically with ${\xi _{\max }}$ = 1.2 and ${\xi _{\min }}$ = 0.4. ${c_1}$ and ${c_2}$ are the individual and global learning factors, respectively. ${r_1}$ and ${r_2}$ are uniform random numbers in the range $\left[ \text{0,1} \right]$ introduced to enhance randomness. Furthermore, ${\bf{\bar g}}_i^*$ and ${{\bf{\bar g}}^*}$ represent the personal best position of the $i$-th particle and the global best position of the entire swarm based on the fitness function, respectively. A sigmoid function is then employed to map the velocity to the interval $\left[ \text{0,1} \right]$, representing the probability that the $i$-th particle takes a value of 1 in the next step, which can be written as
\begin{equation}
    {s}\left( {{\bf{v}}_i^{(j)}(d)} \right) = \frac{1}{{1 + {e^{\left( { - {\bf{v}}_i^{(j)}(d)} \right)}}}}.
    \label{sigmoid}
\end{equation}
Based on the probability, the position of the $i$-th particle in the $j$-th iteration is updated as
\begin{equation}
    {\bf{\bar g}}_i^{(j)}\left( d \right)=
    \begin{cases}
        1, & s\left( {{\bf{v}}_i^{(j)}\left( d \right)} \right) \ge s\left( {{\bf{v}}_i^{(j)}\left( k \right)} \right),\forall k \subseteq \\
        &\left[ {\left( {d \!-\! d{\rm{\% }}M + 1} \right),\left( {d \!+\! M - d{\rm{\% }}M} \right)} \right]\\
        0, & \text{otherwise}
    \end{cases}.
    \label{xuanze}
\end{equation}
During each iteration, the algorithm updates the particle positions according to the fitness function, thereby updating the personal best and global best solutions. Considering the minimum distance constraint for MA, a penalty term is introduced into the objective function of problem (P1). The transformed optimization problem (P2) is rewritten as follows
\allowdisplaybreaks
\begin{subequations}
\label{P1}
\renewcommand{\theequation}{31\alph{equation}} 
\begin{align}
\text{(P2):}\mathop {\max }\limits_{\left\{ {{{\bf{W}}_k}} \right\},{\bf{Z}},{\bf{\Theta }}} &{\cal F}\left( {\left\{ {{{\bf{W}}_k}} \right\}{\rm{,}}{\bf{Z}}{\rm{,}}{\bf{\Theta }}} \right) - {{\cal K}_1}P\left( {{\bf{\tilde t}}_i^{\left( j \right)}} \right),\nonumber\\
\text{s.t.}\quad
& \sum_{k = 1}^K {\rm{Tr}}\left( \mathbf{W}_k \right) + {\rm{Tr}}\left( \mathbf{Z} \right) \le P_{\max}, \\
& \left| \left[ \mathbf{\Theta} \right]_{t,t} \right| = 1,\forall t, \\
&\frac{1}{QL}\sum\limits_{q = 1}^Q \sum\limits_{l = 1}^L \bigg| {\rho _0}{\cal D}\left( {{\theta _l},{\phi _q}} \right) - {{\widehat {\bf{a}}}^H}\left( {{\theta _l},{\phi _q}} \right) \nonumber \\
& \times {\bf{\Theta }}{{\bf{H}}_{dr}}{{\bf{R}}_x}{\bf{H}}_{dr}^H{{\bf{\Theta }}^H}\widehat {\bf{a}}\left( {{\theta _l},{\phi _q}} \right) \bigg|^2 \!\le\! \varepsilon.\\
& \mathbf{W}_k \succeq 0, {\rm{rank}}\left( \mathbf{W}_k \right) = 1,\forall k, \\
& \mathbf{Z} \succeq 0, {\rm{rank}}\left( \mathbf{Z} \right) = 1,
\end{align}
\label{Penalty}
\end{subequations}
\\
where $P\left( {{\bf{\tilde t}}_i^{\left( j \right)}} \right)$ returns the number of MA elements that violate the minimum spacing constraint at position ${\bf{\tilde t}} = {\left[ {{\bf{t}}_{_1}^{\rm{T}}, \ldots ,{\bf{t}}_{_{{N_{\rm{t}}}}}^{\rm{T}}} \right]^{\rm{T}}} \in {\mathbb{C}^{2{N_{\rm{t}}} \times 1}}$, ${{\cal K}_1}$ is the penalty factor. A sufficiently large penalty suppresses infeasible MA position patterns, while an excessively small or large value may respectively weaken constraint enforcement or limit the exploration of high secrecy rate solutions. Therefore, ${\cal K}_1$ is selected to balance feasibility and performance. When MA elements are located at the same discrete candidate position, the penalty term guides the particles towards the feasible region satisfying constraints \eqref{dist1}.

In summary, the BPSO algorithm effectively transforms the MA position selection and minimum distance constraints. As the iterations proceed, the algorithm continuously updates the personal and global best positions while ensuring constraint satisfaction, progressively determining the MA position selection matrix ${{\bf{B}}_{\rm{t}}}$ that meets all requirements.
\subsection{Active Beamforming Design for MA-BS}
In this subsection, we design the beamforming matrix and the AN matrix for the MA-BS. First, Given the fixed RIS phase shift ${\bf{\Theta }}$ and the MA position selection ${{\bf{B}}_{\rm{t}}}$, the optimization problem can be expressed as (P3.0)

\allowdisplaybreaks
\begin{subequations}
\label{P1}
\renewcommand{\theequation}{32\alph{equation}} 
\begin{align}
\text{(P3.0):}\mathop {\max }\limits_{\left\{ {{{\bf{W}}_k}} \right\},{\bf{Z}}} &{\cal F}\left( {\left\{ {{{\bf{W}}_k}} \right\}{\rm{,}}{\bf{Z}}} \right) - {{\cal K}_1}P\left( {{\bf{\tilde t}}_i^{\left( j \right)}} \right),\nonumber\\
\text{s.t.}\quad
& \sum_{k = 1}^K {\rm{Tr}}\left( \mathbf{W}_k \right) + {\rm{Tr}}\left( \mathbf{Z} \right) \le P_{\max}, \\
&\frac{1}{QL}\sum\limits_{q = 1}^Q \sum\limits_{l = 1}^L \bigg| {\rho _0}{\cal D}\left( {{\theta _l},{\phi _q}} \right) - {{\widehat {\bf{a}}}^H}\left( {{\theta _l},{\phi _q}} \right) \nonumber \\
& \times {\bf{\Theta }}{{\bf{H}}_{dr}}{{\bf{R}}_x}{\bf{H}}_{dr}^H{{\bf{\Theta }}^H}\widehat {\bf{a}}\left( {{\theta _l},{\phi _q}} \right) \bigg|^2 \!\le\! \varepsilon.\\
& \mathbf{W}_k \succeq 0, {\rm{rank}}\left( \mathbf{W}_k \right) = 1,\forall k, \\
& \mathbf{Z} \succeq 0, {\rm{rank}}\left( \mathbf{Z} \right) = 1.
\end{align}
\end{subequations}
\\
The objective function is obviously non-convex. Furthermore, the matrix  lifting technique introduces rank-one constraints. Consequently, methods such as SCA, DC programming are employed to address the non-convexity of the optimization problem. First, the objective function is transformed using a first-order Taylor expansion. Let ${\bf{W}}_k^n$ and ${{\bf{Z}}^n}$ denote the values of ${{\bf{W}}_k}$ and ${\bf{Z}}$ at the $n$-th iteration, respectively. By performing the expansion at the point $\left( {{\bf{W}}_k^n,{{\bf{Z}}^n}} \right)$, we obtain
\begin{align}
R_k^{2,l}\left[ {{\bf{W}}_k^n,{{\bf{{\rm Z}}}^n}} \right] 
= &R_k^2\left[ {{\bf{W}}_k^n,{{\bf{{\rm Z}}}^n}} \right] \!+\! {\nabla _{{\bf{W}}_k^n}}R_k^2\left[ {{\bf{W}}_k^n,{{\bf{{\rm Z}}}^n}} \right]  \left( {{{\bf{W}}_k}\! -\! {\bf{W}}_k^n} \right)\nonumber \\
 &+ {\nabla _{{{\bf{{\rm Z}}}^n}}}R_k^2\left[ {{\bf{W}}_k^n,{{\bf{Z}}^n}} \right]  \left( {{\bf{Z}} - {{\bf{ Z}}^n}} \right),
\end{align}
\begin{equation}
    R_k^{{{\bf{W}}_k},{\bf{Z}}} = R_k^1\left[ {{{\bf{W}}_k},{\bf{Z}}} \right] - R_k^{2,l}\left[ {{\bf{W}}_k^n,{{\bf{Z}}^n}} \right],
\end{equation}
where
\begin{equation}
    {\nabla _{{{\bf{Z}}^n}}}R_k^2\left[ {{{\bf{Z}}^n}} \right] = \frac{{{{\bf{H}}_k}}}{{\ln 2\left( {\sum\limits_{c \ne k}^K {{\rm{Tr}}\left( {{{\bf{H}}_k}{{\bf{W}}_c^n}} \right)}  \!+\! {\rm{Tr}}\left( {{{\bf{H}}_k}{{{\bf{Z}}^n}}} \right) \!+\! \sigma _k^2} \right)}},
\end{equation}
\begin{equation}
    {\nabla _{{\bf{W}}_k^n}}R_k^2\left[ {{\bf{W}}_k^n} \right] = 0.
\end{equation}
Similarly,
\begin{align}
C_k^{1,l}\left[ {{\bf{W}}_k^n,{{\bf{Z}}^n}} \right] \!= &C_k^1\left[ {{\bf{W}}_k^n,{{\bf{Z}}^n}} \right] \!+\! {\nabla _{{\bf{W}}_k^n}}C_k^1\left[ {{\bf{W}}_k^n,{{\bf{Z}}^n}} \right]\left( {{{\bf{W}}_k} \!-\! {\bf{W}}_k^n} \right)\nonumber \\
 &+ {\nabla _{{{\bf{Z}}^n}}}C_k^1\left[ {{\bf{W}}_k^n,{{\bf{Z}}^n}} \right]\left( {{\bf{Z}} - {{\bf{Z}}^n}} \right),
\end{align}
\begin{equation}
    C_k^{{{\bf{W}}_k},{\bf{Z}}} = C_k^{1,l}\left[ {{\bf{W}}_k^n,{{\bf{Z}}^n}} \right] - C_k^2\left[ {{{\bf{W}}_k},{\bf{Z}}} \right],
\end{equation}
where
\begin{equation}
    {\nabla _{{{\bf{Z}}^n}}}C_k^1\left[ {{{\bf{Z}}^n}} \right] = \frac{{{{\bf{H}}_e}}}{{\ln 2\left( {{\rm{Tr}}\left( {{{\bf{H}}_e}{{\bf{W}}_k^n}} \right){\rm{ + Tr}}\left( {{{\bf{H}}_e}{{{\bf{Z}}^n}}} \right) + \sigma _e^2} \right)}},
\end{equation}
\begin{equation}
    {\nabla _{{\bf{W}}_k^n}}C_k^1\left[ {{{\bf{W}}_k^n}} \right] = \frac{{{{\bf{H}}_e}}}{{\ln 2\left( {{\rm{Tr}}\left( {{{\bf{H}}_e}{{\bf{W}}_k^n}} \right){\rm{ + Tr}}\left( {{{\bf{H}}_e}{{{\bf{Z}}^n}}} \right) + \sigma _e^2} \right)}}.
\end{equation}
Through the above processing, the objective function is successfully transformed into a concave function
\begin{equation}
    \mathop {\max }\limits_{\left\{ {{{\bf{W}}_k}} \right\},{\bf{Z}}} \sum\limits_{k = 1}^K {\left( {R_k^{{{\bf{W}}_k},{\bf{Z}}} - C_k^{{{\bf{W}}_k},{\bf{Z}}}} \right)}  - {{\cal K}_1}P\left( {{\bf{\tilde t}}_i^{\left( j \right)}} \right).
\end{equation}
For the positive semidefinite matrices $\{{{\bf{W}}_k}\}$, $\forall k$, ${\bf{Z}}$, the two rank-one constraints can be equally expressed as the difference of two convex functions
\begin{equation}
    {\rm{rank}}\left( {{{\bf{W}}_k}} \right) = 1 \Leftrightarrow {\rm{Tr}}\left( {{{\bf{W}}_k}} \right) - {\left\| {{{\bf{W}}_k}} \right\|_2} = 0,\forall k,
\end{equation}
\begin{equation}
    {\rm{rank}}\left( {\bf{Z}} \right) = 1 \Leftrightarrow {\rm{Tr}}\left( {\bf{Z}} \right) - {\left\| {\bf{Z}} \right\|_2} = 0,
\end{equation}
The above equivalence follows from the eigenvalue property of positive semidefinite matrices. Specifically, for $\mathbf{W}_k\succeq 0$, $\mathrm{Tr}(\mathbf{W}_k)$ equals the sum of all eigenvalues, while $\|\mathbf{W}_k\|_2$ equals the largest eigenvalue. Therefore, $\mathrm{Tr}(\mathbf{W}_k)-\|\mathbf{W}_k\|_2=0$ holds if and only if all eigenvalues except the largest one are zero, which indicates that $\mathbf{W}_k$ is rank-one. The same argument also applies to $\mathbf{Z}$. Based on this equivalent transformation, the objective problem can be reformulated as

\begin{align}
\max_{\left\{ \mathbf{W}_k \right\}, \mathbf{Z}} \sum_{k=1}^K &\left( R_k^{\mathbf{W}_k, \mathbf{Z}} - C_k^{\mathbf{W}_k, \mathbf{Z}} - \tau_1 \left( \mathrm{Tr}\left( \mathbf{W}_k \right) - \left\| \mathbf{W}_k \right\|_2 \right)\right) \notag \\  
&- \tau_2 \left( \mathrm{Tr}\left( \mathbf{Z} \right) - \left\| \mathbf{Z} \right\|_2 \right) - \mathcal{K}_1 P\left( \tilde{\mathbf{t}}_i^{(j)} \right),
\end{align}
where ${\tau _1}$ and ${\tau _2}$ are the penalty factors for the rank-one constraints. Since the spectral norm function is convex, it can be lower-bounded by its first-order approximation at the previous iteration point. Specifically, for $\mathbf{W}_k^{l-1}$, we have
\begin{equation}
\|\mathbf{W}_k\|_2
\geq
\|\mathbf{W}_k^{l-1}\|_2
+
\left\langle
\partial\|\mathbf{W}_k^{l-1}\|_2,
\mathbf{W}_k-\mathbf{W}_k^{l-1}
\right\rangle .
\end{equation}
By substituting this lower bound into the penalty term
$-\tau_1(\mathrm{Tr}(\mathbf{W}_k)-\|\mathbf{W}_k\|_2)$ and ignoring the constant terms independent of $\mathbf{W}_k$, we obtain
$-\tau_1
\left\langle
\mathbf{W}_k,
\mathbf{I}-\partial\|\mathbf{W}_k^{l-1}\|_2
\right\rangle ,$ where $\partial {\left\| {{\bf{W}}_k^{l - 1}} \right\|_2}$ denotes the subgradients of the spectral norm at the ($l-1$)-th iteration.
Similarly, the penalty term for $\mathbf{Z}$ is transformed into
$-\tau_2\left\langle \mathbf{Z},\mathbf{I}-\partial\|\mathbf{Z}^{l-1}\|_2\right\rangle$.
This yields the DC-based rank-one approximation used in (P3.1):
\begin{subequations}
\label{P3.1}
\renewcommand{\theequation}{46\alph{equation}}
\begin{align}
    \text{(P3.1):}\quad 
    \mathop{\max}\limits_{\{\mathbf{W}_k\},\mathbf{Z}}
    &\sum_{k=1}^K \Bigl(R_k^{\mathbf{W}_k,\mathbf{Z}} - C_k^{\mathbf{W}_k,\mathbf{Z}}\notag \\
    &- \tau_1 \bigl\langle \mathbf{W}_k, \mathbf{I} - \partial \|\mathbf{W}_k^{l-1}\|_2 \bigr\rangle\Bigr) \notag\\
    & - \tau_2 \bigl\langle \mathbf{Z}, \mathbf{I} - \partial \|\mathbf{Z}^{l-1}\|_2 \bigr\rangle
    - \mathcal{K}_1 P\bigl(\tilde{\mathbf{t}}_i^{(j)}\bigr),
    \label{eq:p2_obj}\notag \\
    \text{s.t.}\quad
    &\sum_{k=1}^K \operatorname{Tr}(\mathbf{W}_k) + \operatorname{Tr}(\mathbf{Z}) \le P_{\max},\\
    &\frac{1}{QL}\sum_{q=1}^Q \sum_{l=1}^L \Bigl| \rho_0 \mathcal{D}(\theta_l,\phi_q)
    - \widehat{\mathbf{a}}^H(\theta_l,\phi_q) \notag \\
    &\times\mathbf{\Theta} \mathbf{H}_{\rm dr} \mathbf{R}_x \mathbf{H}_{\rm dr}^H \mathbf{\Theta}^H \widehat{\mathbf{a}}(\theta_l,\phi_q) \Bigr|^2 \le \varepsilon, \\
    &\mathbf{W}_k \succeq 0,\;\forall k,\\
    &\mathbf{Z} \succeq 0, \end{align}
\end{subequations}
For the positive semidefinite matrices $\{{{\bf{W}}_k}\},\forall k$ and $\bf{Z}$, these subgradients can be computed as ${{\bf{w}}_{k1}}{\bf{w}}_{k1}^{\rm{H}}$ and ${{\bf{z}}_1}{\bf{z}}_1^H$, where ${{\bf{w}}_{k1}} \in {\mathbb{C}^{{N_{\rm{t}}} \times 1}}$ and ${{\bf{z}}_1} \in {\mathbb{C}^{{N_{\rm{t}}} \times 1}}$ are the eigenvectors corresponding to the largest singular values of their respective matrices\cite{robust}. Through this transformation, the optimization problem has been successfully reformulated into a convex problem, which can be solved using CVX\cite{cvx}. 
\subsection{Passive Beamforming Design for RIS}
When the beamforming and AN design of the MA-BS $\left\{ {{{\bf{W}}_k}} \right\},\forall k$, $\bf{Z}$ and the MA position selection ${{\bf{B}}_{\rm{t}}}$ are given, the optimization problem can be reformulated as

\allowdisplaybreaks
\begin{subequations}
\label{P1}
\renewcommand{\theequation}{47\alph{equation}} 
\begin{align}
\text{(P4.0):}\mathop {\max }\limits_{{\bf{\Theta }}} \quad&{\cal F}\left( {{\bf{\Theta }}} \right) - {{\cal K}_1}P\left( {{\bf{\tilde t}}_i^{\left( j \right)}} \right),\nonumber\\
\text{s.t.}\quad
& \left| \left[ \mathbf{\Theta} \right]_{t,t} \right| = 1,\forall t, \\
&\frac{1}{QL}\sum\limits_{q = 1}^Q \sum\limits_{l = 1}^L \bigg| {\rho _0}{\cal D}\left( {{\theta _l},{\phi _q}} \right) - {{\widehat {\bf{a}}}^H}\left( {{\theta _l},{\phi _q}} \right) \nonumber \\
& \times {\bf{\Theta }}{{\bf{H}}_{dr}}{{\bf{R}}_x}{\bf{H}}_{dr}^H{{\bf{\Theta }}^H}\widehat {\bf{a}}\left( {{\theta _l},{\phi _q}} \right) \bigg|^2 \!\le\! \varepsilon.
\end{align}
\end{subequations}
\\
To handle the highly non-convex optimization problem, we first set ${\bf{o}} = {\left[ {{e^{j{\alpha _1}}}, \ldots ,{e^{j{\alpha _t}}}, \ldots ,{e^{j{\alpha _T}}}} \right]^T}$ and introduce auxiliary variable ${\bf{O}} = {\bf{o}}{{\bf{o}}^H} \in {\mathbb{C}^{T \times T}}$. The optimization problem can be reconstructed as (P4.1), 
\begin{figure*}[t]
\centering
\begin{subequations}
\label{P3.1}
\renewcommand{\theequation}{48\alph{equation}}
\begin{align}
    \text{(P4.1):}
    \mathop {\max }\limits_{\bf{O}} &\sum\limits_{k = 1}^K \begin{array}{l}
\left( \begin{array}{l}
\underbrace {{{\log }_2}\left( {\sum\limits_{c = 1}^K {{\rm{Tr}}\left( {{{\bf{R}}_{kc}}{\bf{O}}} \right)}  \!+\! {\rm{Tr}}\left( {{{\bf{R}}_{kz}}{\bf{O}}} \right) \!+\! \sigma _k^2} \right)}_{R_k^3} \!- 
\underbrace {{{\log }_2}\left( {\sum\limits_{c \ne k}^K {{\rm{Tr}}\left( {{{\bf{R}}_{kc}}{\bf{O}}} \right)}  \!+\! {\rm{Tr}}\left( {{{\bf{R}}_{kz}}{\bf{O}}} \right) \!+\! \sigma _k^2} \right)}_{R_k^4} \!- \\
\underbrace {{{\log }_2}\left( {{\rm{Tr}}\left( {{{\bf{R}}_{ek}}{\bf{O}}} \right) + {\rm{Tr}}\left( {{{\bf{R}}_{ez}}{\bf{O}}} \right) + \sigma _e^2} \right)}_{C_k^3} + 
\underbrace {{{\log }_2}\left( {{\rm{Tr}}\left( {{{\bf{R}}_{ez}}{\bf{O}}} \right) + \sigma _e^2} \right)}_{C_k^4}
\end{array} \right)\\
 - {\tau _3}\left\langle {{\bf{O}},{\bf{I}} - \partial {{\left\| {{{\bf{O}}^{l - 1}}} \right\|}_2}} \right\rangle  - {{\cal K}_1}P\left( {{\bf{\tilde t}}_i^{\left( j \right)}} \right),
\end{array} \nonumber\\
    \text{s.t.}\quad
    &{{\bf{O}}_{t,t}} = 1,\forall t,\\
    &\frac{1}{QL}\sum_{q=1}^Q \sum_{l=1}^L \Bigl| \rho_0 \mathcal{D}(\theta_l,\phi_q)
    -\widehat{\mathbf{a}}^H(\theta_l,\phi_q)\mathbf{\Theta} \mathbf{H}_{\rm dr} \mathbf{R}_x \mathbf{H}_{\rm dr}^H \mathbf{\Theta}^H \widehat{\mathbf{a}}(\theta_l,\phi_q) \Bigr|^2 \le \varepsilon, \\
    &{\rm{rank}}\left( {\bf{O}} \right) = 1, \mathbf{O} \succeq 0,
\end{align}
\end{subequations}
\hrule
\end{figure*}
where ${\tau _3}$ is the penalty factor for the rank-one constraint, and the relevant matrices are defined as follows\cite{1}
\begin{equation}
    {{\bf{R}}_{kc}} = \left( {{{\bf{h}}_{rk}}{\bf{h}}_{rk}^H} \right) \odot {\left( {{{\bf{H}}_{dr}}{{\bf{W}}_c}{\bf{H}}_{dr}^H} \right)^T},
\end{equation}
\begin{equation}
    {{\bf{R}}_{kz}} = \left( {{{\bf{h}}_{rk}}{\bf{h}}_{rk}^H} \right) \odot {\left( {{{\bf{H}}_{dr}}{\bf{ZH}}_{dr}^H} \right)^T},
\end{equation}
\begin{equation}
    {{\bf{R}}_{ek}} = \left( {{{\bf{h}}_{re}}{\bf{h}}_{re}^H} \right) \odot {\left( {{{\bf{H}}_{dr}}{{\bf{W}}_c}{\bf{H}}_{dr}^H} \right)^T},
\end{equation}
\begin{equation}
    {{\bf{R}}_{ez}} = \left( {{{\bf{h}}_{re}}{\bf{h}}_{re}^H} \right) \odot {\left( {{{\bf{H}}_{dr}}{\bf{ZH}}_{dr}^H} \right)^T}.
\end{equation}

Additionally, to address the non-convex sensing constraint, the following definition is introduced:
\begin{equation}
    {{\bf{H}}_{\rm{r}}} \buildrel \Delta \over = {\rm{diag}}\left( {{{\widehat {\bf{a}}}^H}\left( {{\theta _l},{\phi _q}} \right)} \right){{\bf{H}}_{dr}}{{\bf{R}}_x}{\bf{H}}_{dr}^H{\rm{diag}}\left( {\widehat {\bf{a}}\left( {{\theta _l},{\phi _q}} \right)} \right).
\end{equation}
The gain of the actual beampattern in the direction $\left( {{\theta _l},{\phi _q}} \right)$ can be rewritten as
\begin{equation}
    {\cal P}\left( {{\theta _l},{\phi _q}} \right) = {{\bf{o}}^T}{{\bf{H}}_{\rm{r}}}{{\bf{o}}^*}.
\end{equation}
Based on the matrix trace operation, the sensing constraint can be reconstructed into the following convex constraint form:
\begin{equation}
    \frac{1}{Q}\frac{1}{L}\sum\limits_{q = 1}^Q {\sum\limits_{l = 1}^L {{{\left| {{\rho _0}{\cal D}\left( {{\theta _l},{\phi _q}} \right) - {\rm{Tr}}\left( {{{\bf{H}}_r}{{\bf{O}}^*}} \right)} \right|}^2}} }  \le \varepsilon.
\end{equation}
The DC programming is reused to handle the rank-one constraint in the optimization problem  (P4.1), which can be equivalent to the difference between the following two convex functions:
\begin{equation}
    {\rm{rank}}\left( {\bf{O}} \right) = 1 \Leftrightarrow {\rm{Tr}}\left( {\bf{O}} \right) - {\left\| {\bf{O}} \right\|_2} = 0.
\end{equation}
After adding a penalty term and transforming the matrix spectral norm, the optimization can be expressed as
\begin{subequations}
\label{P4.2}
\renewcommand{\theequation}{57\alph{equation}}
\begin{align}
\text{(P4.2):}\mathop {\max }\limits_{\bf{O}} \sum\limits_{k = 1}^K &{\left( {R_k^3 - R_k^4 - C_k^3 + C_k^4} \right)}  -\nonumber\\ 
&{\tau _3}\left\langle {{\bf{O}},{\bf{I}} - \partial {{\left\| {{{\bf{O}}^{l - 1}}} \right\|}_2}} \right\rangle  - {{\cal K}_1}P\left( {{\bf{\tilde t}}_i^{\left( j \right)}} \right),\nonumber\\
    \text{s.t.}\quad
    &{{\bf{O}}_{t,t}} = 1,\forall t,\\
    &\frac{1}{QL}\sum\limits_{q = 1}^Q {\sum\limits_{l = 1}^L {{{\left| {{\rho _0}{\cal D}\left( {{\theta _l},{\phi _q}} \right) \!-\! {\rm{Tr}}\left( {{{\bf{H}}_r}{{\bf{O}}^*}} \right)} \right|}^2}} }  \!\le\! \varepsilon ,\\
    &\mathbf{O} \succeq 0.
\end{align}
\end{subequations}
Though the rank-one constraint has been relaxed, the objective function is still non-convex. SCA is adopted again to perform a convex approximation of the objective function. Let the value of $\bf{O}$ at the $n$-th iteration be ${{\bf{O}}^n}$, and we obtain
\begin{equation}
    R_k^{4,l}\left[ {{{\bf{O}}^n}} \right] = R_k^4\left[ {{{\bf{O}}^n}} \right] + {\nabla _{{{\bf{O}}^n}}}R_k^4\left[ {{{\bf{O}}^n}} \right] \left( {{\bf{O}} - {{\bf{O}}^n}} \right),
\end{equation}
\begin{equation}
    R_k^{\bf{O}} = R_k^3\left[ {\bf{O}} \right] - R_k^{4,l}\left[ {{{\bf{O}}^n}} \right],
\end{equation}
\begin{equation}
    C_k^{3,l}\left[ {{{\bf{O}}^n}} \right] = C_k^3\left[ {{{\bf{O}}^n}} \right] + {\nabla _{{{\bf{O}}^n}}}C_k^3\left[ {{{\bf{O}}^n}} \right] \left( {{\bf{O}} - {{\bf{O}}^n}} \right),
\end{equation}
\begin{equation}
    C_k^{\bf{O}} = C_k^{3,l}\left[ {{{\bf{O}}^n}} \right] - C_k^4\left[ {{{\bf{O}}^n}} \right],
\end{equation}
where
\begin{equation}
    {\nabla _{{{\bf{O}}^n}}}R_k^4\left[ {{{\bf{O}}^n}} \right] = \frac{{\sum\limits_{c \ne k}^K {{{\bf{R}}_{kc}}}  + {{\bf{R}}_{kz}}}}{{\ln 2\left( {\sum\limits_{c \ne k}^K {{\rm{Tr}}\left( {{{\bf{R}}_{kc}}{{\bf{O}}^n}} \right)}  \!+\! {\rm{Tr}}\left( {{{\bf{R}}_{kz}}{{\bf{O}}^n}} \right) \!+\! \sigma _k^2} \right)}},
\end{equation}
\begin{equation}
    {\nabla _{{{\bf{O}}^n}}}C_k^3\left[ {{{\bf{O}}^n}} \right] = \frac{{{{\bf{R}}_{ek}} + {{\bf{R}}_{ez}}}}{{\ln 2\left( {{\rm{Tr}}\left( {{{\bf{R}}_{ek}}{{\bf{O}}^n}} \right) + {\rm{Tr}}\left( {{{\bf{R}}_{ez}}{{\bf{O}}^n}} \right) + \sigma _e^2} \right)}}.
\end{equation}
Through the above operations, the objective function is transformed into a concave function form, and the problem is finally converted into problem (P4.3) as follows
\begin{subequations}
\label{P4.3}
\renewcommand{\theequation}{64\alph{equation}}
\begin{align}
\text{(P4.3):}\mathop {\max }\limits_{\bf{O}} &\sum\limits_{k = 1}^K {\left( {R_k^{\bf{O}} - C_k^{\bf{O}}} \right)}  - {\tau _3}\left\langle {{\bf{O}},{\bf{I}} - \partial {{\left\| {{{\bf{O}}^{r - 1}}} \right\|}_2}} \right\rangle  \nonumber\\
&- {{\cal K}_1}P\left( {{\bf{\tilde t}}_i^{\left( j \right)}} \right),\nonumber\\
    \text{s.t.}\quad
    &{{\bf{O}}_{t,t}} = 1,\forall t,\\
    &\frac{1}{QL}\sum\limits_{q = 1}^Q {\sum\limits_{l = 1}^L {{{\left| {{\rho _0}{\cal D}\left( {{\theta _l},{\phi _q}} \right) \!-\! {\rm{Tr}}\left( {{{\bf{H}}_r}{{\bf{O}}^*}} \right)} \right|}^2}} }  \!\le\! \varepsilon ,\\
    &\mathbf{O} \succeq 0.
\end{align}
\end{subequations}
In summary, the processes above solve the non-convex objective function and constraints. We successfully transform the optimization problem into a convex sub-problem that can be effectively solved using standard convex optimization solvers like CVX.

\begin{algorithm}[!t]
\renewcommand{\algorithmicrequire}{\textbf{Input:}}
\renewcommand{\algorithmicensure}{\textbf{Output:}}
\caption{Joint Discrete Antenna Positioning and Beamforming Design Optimization Algorithm}
\label{alg:joint_optimization}
\begin{algorithmic}[1]
    \REQUIRE $\mathbf{B}_{\rm{t}}^{(0)}$, $\{\mathbf{W}_k^{(0)}\}$, $\mathbf{Z}^{(0)}$, $\mathbf{\Theta}^{(0)}$, maximum iteration number $U_{\max}$, convergence threshold $\eta$, and iteration index $u=0$.
    
    \REPEAT
        \STATE \textbf{Sub-problem 1: MA Position Optimization} \\
        Given  $\{\mathbf{W}_k^{(u-1)}\}$, $\mathbf{Z}^{(u-1)}$, and $\mathbf{\Theta}^{(u-1)}$, calculate the fitness function value to obtain the local optimal $\tilde{\mathbf{b}}_i^*$ and the current global optimal $\tilde{\mathbf{b}}^*$. Update the velocities of $i$-th particle $\mathbf{v}_i^{(j)}(d)$ and solve \eqref{sigmoid}-\eqref{xuanze} to obtain $\mathbf{B}_{\rm{t}}^{(u)}$.

        \STATE \textbf{Sub-problem 2: Active Beamforming Design for MA-BS} \\
        Given $\mathbf{B}_{\rm{t}}^{(u)}$ and $\mathbf{\Theta}^{(u-1)}$, calculate the equivalent channels. Solve problem (P3.0) by applying SCA and DC programming techniques to obtain $\{\mathbf{W}_k^{(u)}\}$ and $\mathbf{Z}^{(u)}$.

        \STATE \textbf{Sub-problem 3: Passive Beamforming Design for RIS} \\
        Given $\mathbf{B}_{\rm{t}}^{(u)}$, $\{\mathbf{W}_k^{(u)}\}$, and $\mathbf{Z}^{(u)}$, solve problem (P4.0) by applying SCA and DC programming techniques to obtain the auxiliary variable $\mathbf{O}^{(u)}$. Recover $\mathbf{\Theta}^{(u)}$ by extracting the principal eigenvector and projecting it onto the unit modulus constraint.
        
        \STATE \text{Update} the iteration index $u = u + 1$.
        \STATE Calculate the current secrecy rate $R_{\rm{sec}}^{(u)} = \sum_{k=1}^K D_k^{(u)}$.
        
    \UNTIL{ $|R_{\rm{sec}}^{(u)} - R_{\rm{sec}}^{(u-1)}| \le \eta $ or $u \ge U_{\max}$. }
\ENSURE $\mathbf{B}_{\rm{t}}$, $\{\mathbf{W}_k\}$, $\mathbf{Z}$, $\mathbf{\Theta}$ and total secrecy rate $R_{\rm{sec}}$.
\end{algorithmic}
\end{algorithm}
\subsection{Convergence and Computational Complexity Analysis}

In this subsection, we analyze the convergence and computational complexity of the proposed algorithm. The proposed joint discrete antenna positioning and beamforming design optimization algorithm is summarized in \textbf{Algorithm} \ref{alg:joint_optimization}. To facilitate understanding, the proposed algorithm is briefly explained before the pseudocode. Since the MA position selection, active beamforming and AN design, and RIS phase shifts are highly coupled, problem (P1) is solved using an AO framework. The BPSO module first determines the MA positions. Then, with the MA positions fixed, the active beamforming and AN covariance are optimized. Finally, the RIS phase shifts are updated to reshape the reflected channel and satisfy the sensing constraint. These steps are repeated until the secrecy rate converges.

\subsubsection{Convergence Analysis} The convergence of the proposed AO-based optimization algorithm is guaranteed by the non-decreasing property of the objective function value across iterations. Specifically, in the MA position selection optimization step, the BPSO algorithm is employed. The global best position of the particles is updated only if the new position yields a higher fitness value than the current best. The pattern ensures that the objective function value is non-decreasing in this sub-problem. For the active and passive beamforming sub-problems, we utilize the SCA method. In each SCA iteration, the non-convex objective function is approximated by a concave lower bound derived from the first-order Taylor expansion. Maximizing this alternative function guarantees that the original objective function value does not decrease after each update. Furthermore, due to the practical transmit power constraint ${P_{\max }}$, the achievable secrecy rate of the system is upper-bounded. In conclusion, since the objective function is non-decreasing and bounded from above, the proposed algorithm is guaranteed to converge to a set of stable solutions. 
\subsubsection{Computational Complexity Analysis}

The total computational complexity of \textbf{Algorithm} \ref{alg:joint_optimization} mainly depends on the number of AO iterations $U_{\rm max}$ and the complexity of solving the three sub-problems in each iteration. For the MA position optimization sub-problem, the dominant cost lies in the fitness evaluation of all particles. For each particle, the equivalent channels need to be updated according to the selected MA positions, and the secrecy rate related terms need to be evaluated for all legitimate users. Therefore, the complexity of the BPSO-based MA position optimization is approximately given by
$O\left(I_{\rm BPSO}N_p\left((K+1)TN_{\rm{t}}+K^2N_{\rm{t}}+QLTN_{\rm{t}}\right)\right)$,
where the first term corresponds to the equivalent channel computation, the second term is associated with the multi-user secrecy rate evaluation, and the last term is incurred when the sensing beampattern MSE constraint is evaluated.
The second sub-problem involves the optimization of active beamforming. According to the complexity analysis of the interior-point method, the complexity is approximately ${\cal O}\left( {{I_{\rm{s}}}{{\left( {K{N_{\rm{t}}}} \right)}^{3.5}}} \right)$, where ${I_{\rm{s}}}$ is the number of SCA iterations. Similarly, RIS phase shift design optimizes a $T \times T$ matrix variable using SCA and DC programming. The complexity is on the order of ${\cal O}\left( {{I_{\rm{s}}}{T^{3.5}}} \right)$. Thus, the overall computational complexity of Algorithm~1 can be expressed as
$\mathcal{O}\bigl(
U_{\max}\bigl[
I_{\rm BPSO}N_p
\bigl((K+1)TN_{\rm t}+K^2N_{\rm t}+QLTN_{\rm t}\bigr)
+\allowbreak
I_s(KN_{\rm t})^{3.5}
+\allowbreak
I_sT^{3.5}
\bigr]
\bigr)$.

\section{Numerical Results}\label{sec4}
In this section, we present numerical results to demonstrate the convergence behavior and to validate the performance of the proposed joint discrete antenna positioning and beamforming design optimization algorithm in the RIS-assisted MA secure ISAC system. Unless otherwise specified, the specific parameters are set as follows: The MA-BS and RIS are located at coordinates (0 m, 0 m) and (12 m, 12 m), respectively. The $K$ legitimate users are distributed within a circular area centered at (16 m, 16 m) with a radius of 5 m. The same representative user-location realization is adopted in Figs.~2-9. The actual position of the eavesdropper is (13 m, 20 m), posing a high security threat to the system. It should be noted that the adopted coordinates represent a compact urban blockage scenario. Since the passive RIS does not contain an active power amplifier, it cannot compensate
for the cascaded path loss over the BS-RIS and RIS-user links \cite{r2c3}. Therefore, the 20~m-level geometry is adopted to maintain a meaningful reflected-link budget.   The carrier frequency is set to 5 GHz to represent a typical sub-6 GHz wireless communication scenario \cite{r2c2}, corresponding to a wavelength of approximately $\lambda=$ 0.06 m. It is worth noting that the proposed framework can be applied to other carrier frequencies by adjusting the wavelength-dependent parameters. The remaining parameters are listed in \textbf{Table} \ref{parameters}.

\begin{table}[t]
\centering
\caption{Simulation Parameters}
\label{parameters}

\begin{tabular}{
    c
    l
    c
}
\toprule
\textbf{Parameters}
& \textbf{Description}
& \textbf{Values} \\
\midrule

$\lambda$
& Wavelength
& 0.06 m \\

$P_{\rm max}$
& The MA-BS power
& 25 dBm \\

$M$
& Number of MA candidate positions
& 16 \\

$N_{\rm t}$
& Number of MA elements
& 4 \\

$K$
& Number of legitimate users
& 4 \\

$D_{\min}$
& Minimum distance constraint of MA
& 0.03 m \\

$n_k$
& AWGN power at legitimate users
& $-90$ dBm \\

$n_e$
& AWGN power at eavesdropper
& $-90$ dBm \\

$\varpi$
& Path loss exponent of user channel
& 2.7 \\

$\iota$
& Path loss exponent of eavesdropper channel
& 2.7 \\

$C_0$
& Path loss at the reference distance
& $-25$ dB \\

$\alpha$
& Channel fading factor from MA-BS to RIS
& $-60$ dB \\

$L$
& Number of directions in elevation domain
& 6 \\

$Q$
& Number of directions in azimuth domain
& 6 \\

$I$
& Number of particles
& 30 \\

$\tau_1$
& Penalty factor for $\{\mathbf{W}_k\}$ rank-one constraints
& 100 \\

$\tau_2$
& Penalty factor for $\mathbf{Z}$ rank-one constraint
& 100 \\

$\tau_3$
& Penalty factor for $\mathbf{O}$ rank-one constraint
& 100 \\

$U_{\max}$
& Number of AO iterations
& 30 \\

$N_p$
& Number of BPSO particles
& 30 \\

$I_{\rm BPSO}$
& Number of BPSO iterations
& 30 \\

$\xi_{\max},\,\xi_{\min}$
& Maximum and minimum inertia weights
& 1.2, 0.4 \\

$c_1,\,c_2$
& Individual and global learning factors
& 1.8, 1.8 \\

${\cal K}_1$
& Penalty factor for MA distance constraint
& $10^3$ \\

\bottomrule
\end{tabular}

\end{table}

\vspace{-0.7em}
\begin{figure}[H]
    \centering
    \includegraphics[
        width=0.89\linewidth,
        height=0.57\linewidth
    ]{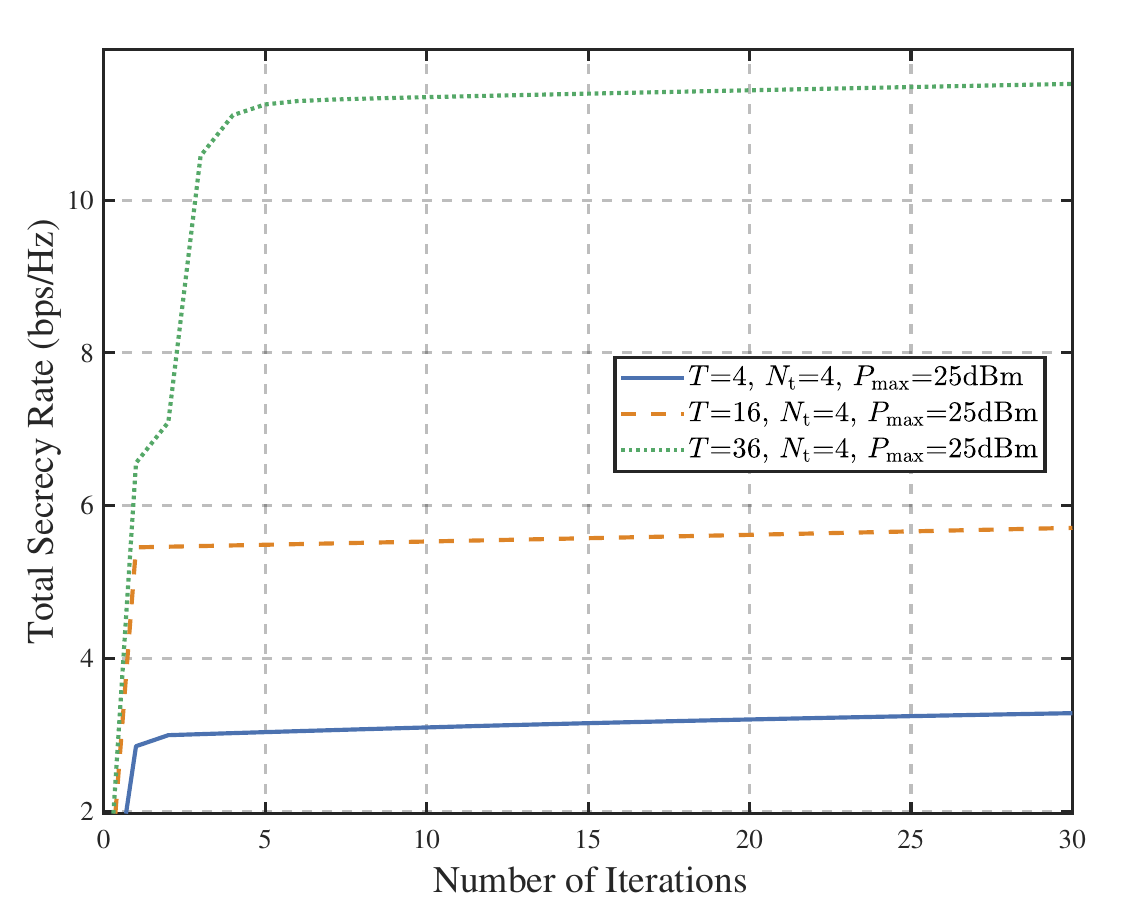}
    \caption{Convergence behavior of the proposed algorithm.}
    \label{fig:2}
    \vspace{1.5em}
\end{figure}

\vspace{-0.2em}

First, the convergence behavior of the proposed algorithm is investigated. Fig. \ref{fig:2} illustrates the trend of the system's total secrecy rate versus the number of iterations under different numbers of RIS reflecting elements. It can be observed that the secrecy rate gradually increases with the number of iterations and eventually converges. Furthermore, by comparing the cases where the number of RIS reflecting elements is 4, 16 and 36, it is evident that the secure communication performance improves as the number of elements increases. This indicates that increasing the number of RIS reflecting elements provides higher spatial DoFs for the system's beamforming design. Thereby the secure communication performance is enhanced.

The proposed joint optimization algorithm is compared with the following ablation and benchmark schemes. 1) Fixed MA, where the MA positions are randomly selected and remain fixed, while the active beamforming, AN covariance, and RIS phase shifts are optimized. This scheme isolates the contribution of discrete MA position optimization. 2) Random RIS, where the RIS phase shifts are randomly generated and fixed, while the MA positions, active beamforming, and AN covariance are optimized. This scheme evaluates the contribution of RIS passive beamforming optimization. 3) Without AN, where no AN is transmitted, while the MA positions and the active and passive beamforming variables are jointly optimized. This scheme quantifies the secrecy gain provided by AN-aided eavesdropping suppression. 4) Greedy-AO, where the MA elements are sequentially updated by enumerating their feasible candidate positions in a one-round greedy search. The selected MA positions are then fixed, while the active beamforming, AN covariance, and RIS phase shifts are optimized using the same AO procedure as in the proposed scheme. This benchmark evaluates the advantage of the proposed BPSO-based discrete position search over a lower-complexity greedy strategy. 5) RIS Impairments, where the proposed joint optimization is first performed under ideal RIS conditions, after which bounded phase errors are imposed on the optimized RIS coefficients during performance evaluation. This scheme examines the robustness of the proposed design against practical RIS hardware imperfections. These comparisons isolate the effects of MA position optimization, RIS phase optimization, AN design, and the adopted discrete positioning strategy, while also evaluating the robustness of their combined gains.

\vspace{-0.5em}
\begin{figure}[H]  
    \centering
\includegraphics[width=0.86\linewidth]{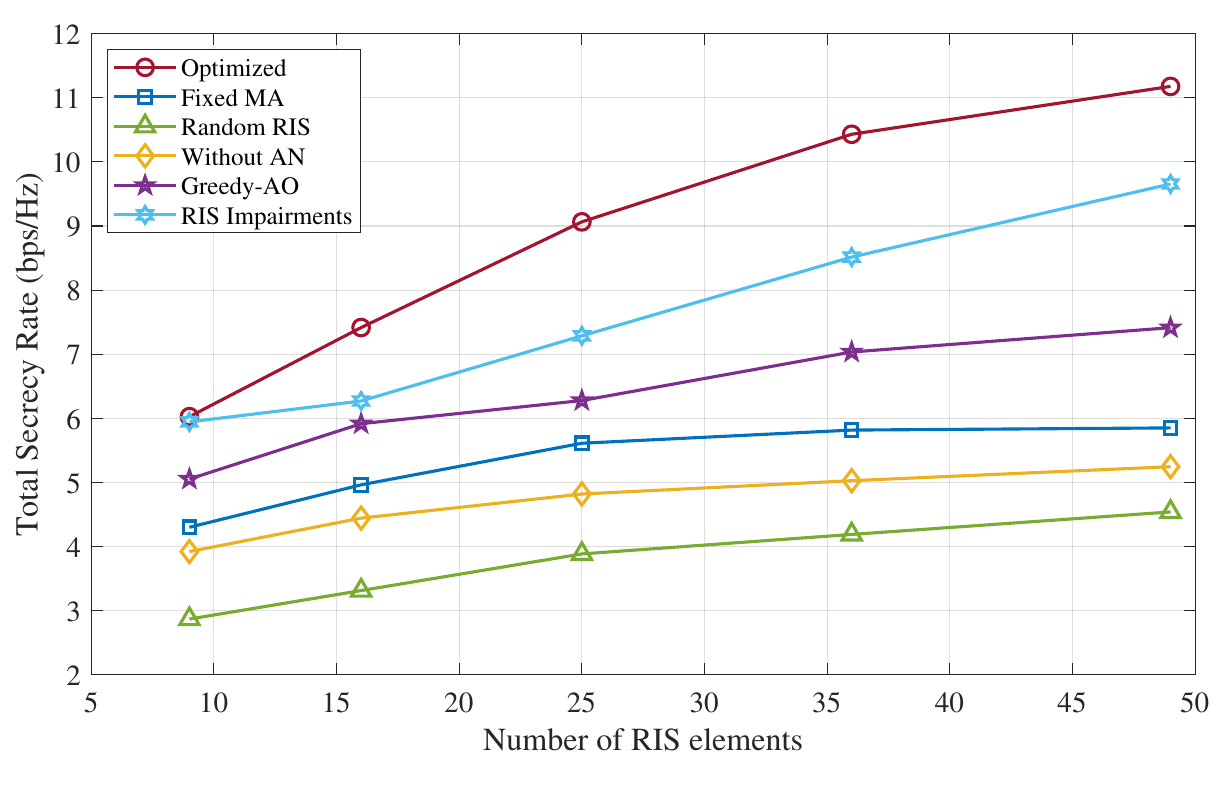} 
\vspace{-1em}
    \caption{\ Total secrecy rate versus the number of RIS elements.}
    \label{fig:3}
    \vspace{2em}
\end{figure}

Fig. \ref{fig:3} illustrates the total secrecy rate versus the number of RIS reflecting elements. Increasing the number of RIS elements enlarges the effective reflecting aperture and provides more passive beamforming degrees of freedom, enabling the reflected signals to be coherently strengthened toward the legitimate users while being suppressed toward the eavesdropper. Therefore, the proposed scheme achieves a considerable secrecy rate improvement as the RIS size increases. In contrast, Random RIS cannot fully exploit the additional reflecting elements because the randomly configured phases do not guarantee coherent signal combining. The performance gap caused by RIS Impairments also becomes more evident for a larger RIS, since phase errors reduce the coherent array gain that would otherwise be obtained from the increased number of reflecting elements.

\vspace{-0.5em}
\begin{figure}[H]  
    \centering
\includegraphics[width=0.85\linewidth]{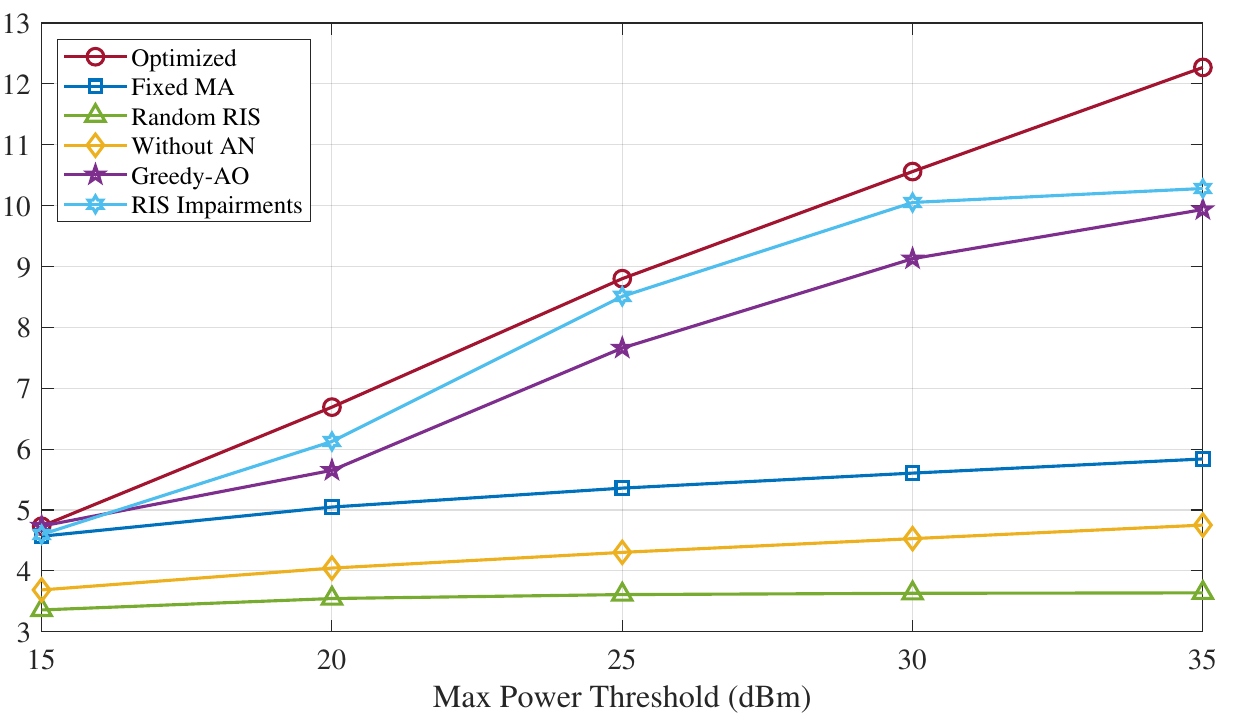} 
    \caption{\ Total secrecy rate versus maximum power threshold.}
    \label{fig:4}
    \vspace{1.5em}
\end{figure}

Fig. \ref{fig:4} shows the effect of the maximum transmit power on the total secrecy rate. A larger power budget provides more resources for information transmission and AN generation, resulting in a substantial secrecy rate improvement for the proposed scheme. Nevertheless, increasing the transmit power alone cannot remove the limitations caused by fixed MA positions, random RIS phases, or the absence of AN, because the additional power may also enhance multiuser interference or the signal received by the eavesdropper. Consequently, these schemes exhibit much slower improvements or gradually approach saturation. The results indicate that the secrecy gain at high transmit power mainly depends on the effective coordination of spatial configuration, beamforming, and AN rather than on power enhancement alone.

\vspace{-0.5em}
\begin{figure}[H]  
    \centering
\includegraphics[width=0.88\linewidth]{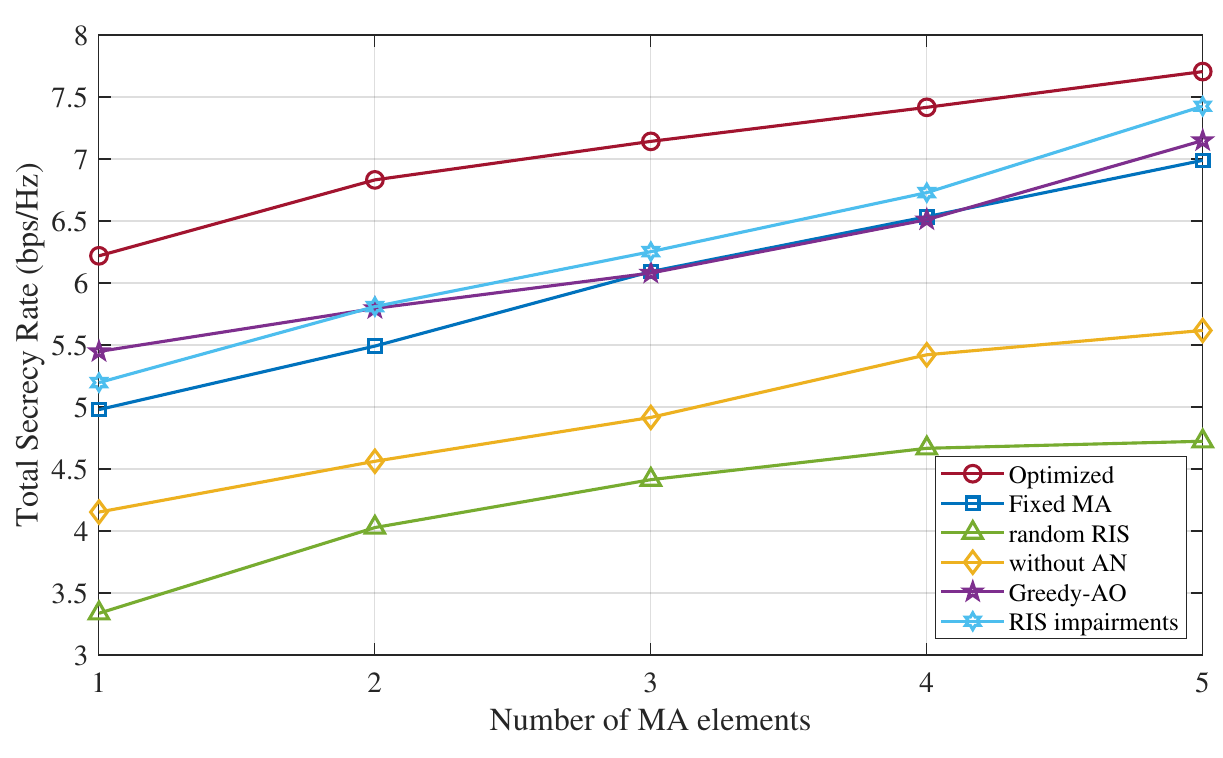} 
\vspace{-1.1em}
    \caption{\ Total secrecy rate versus the number of MA elements.}
    \label{fig:5}
    \vspace{2em}
\end{figure}

Fig.~\ref{fig:5} presents the total secrecy rate versus the number of MA elements. Adding more MA elements increases the active array gain and provides additional spatial degrees of freedom for enhancing the legitimate channels, suppressing multiuser interference, and shaping AN toward the eavesdropper. The proposed scheme consistently achieves the highest secrecy rate because the positions of all MA elements are jointly optimized with the active and passive beamforming variables. However, the performance improvement becomes more moderate as the number of MA elements increases, since the available transmit power is shared among more spatial dimensions and the overall performance is also constrained by the finite RIS aperture. These results confirm that the gain provided by additional MA elements can be effectively exploited only when their positions and the remaining transmission variables are properly coordinated.

\vspace{-0.5em}
\begin{figure}[H]  
    \centering
\includegraphics[width=0.95\linewidth]{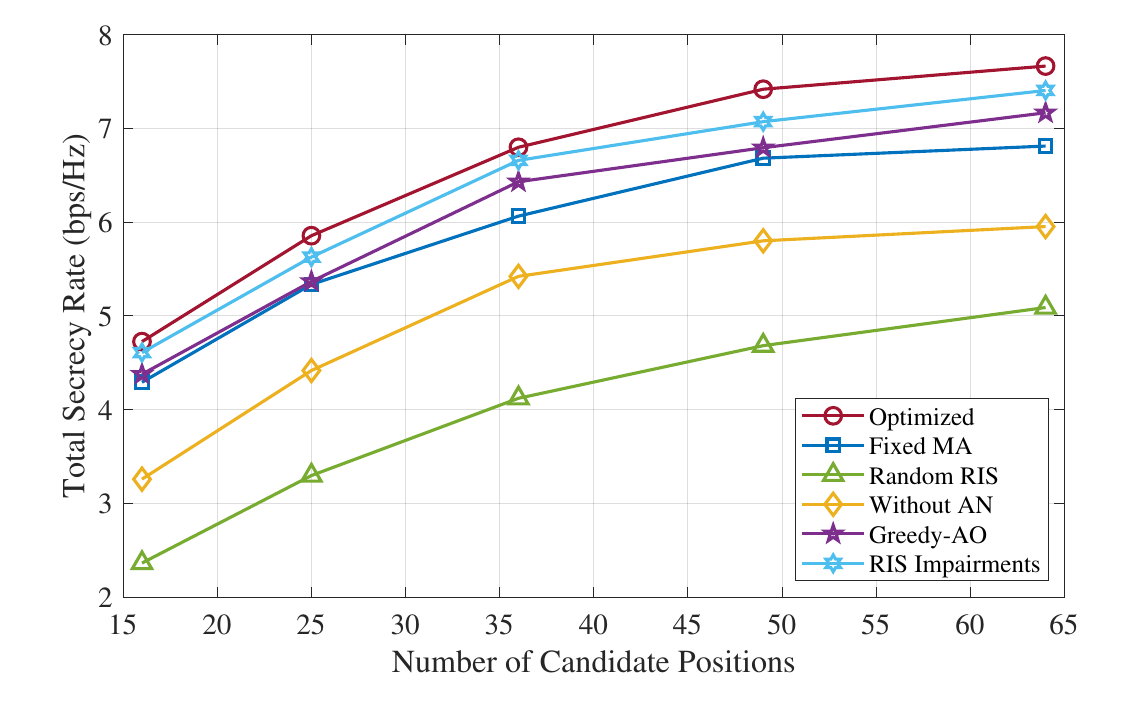} 
\vspace{-0.8em}
    \caption{\ Total secrecy rate versus the number of candidate positions.}
    \label{fig:6}
    \vspace{1.7em}
\end{figure}
\vspace{-0.3em}

Fig.~\ref{fig:6} demonstrates the impact of the number of candidate positions on the system's security performance. It can be observed that the total secrecy rate generally increases as the number of MA candidate positions grows. This is because a larger candidate set provides more spatial configurations for reshaping the cascaded MA-RIS-user and MA-RIS-eavesdropper channels. By selecting favorable discrete positions, the MA-BS can enhance the legitimate links while reducing information leakage. Therefore, the proposed scheme achieves a higher secrecy rate than Fixed MA and Greedy-AO by more effectively exploiting the enlarged discrete positioning space. The remaining performance gap under RIS Impairments is mainly caused by the loss of coherent reflection gain due to phase shift errors.

\begin{figure}[H]  
    \centering
\includegraphics[width=0.86\linewidth]{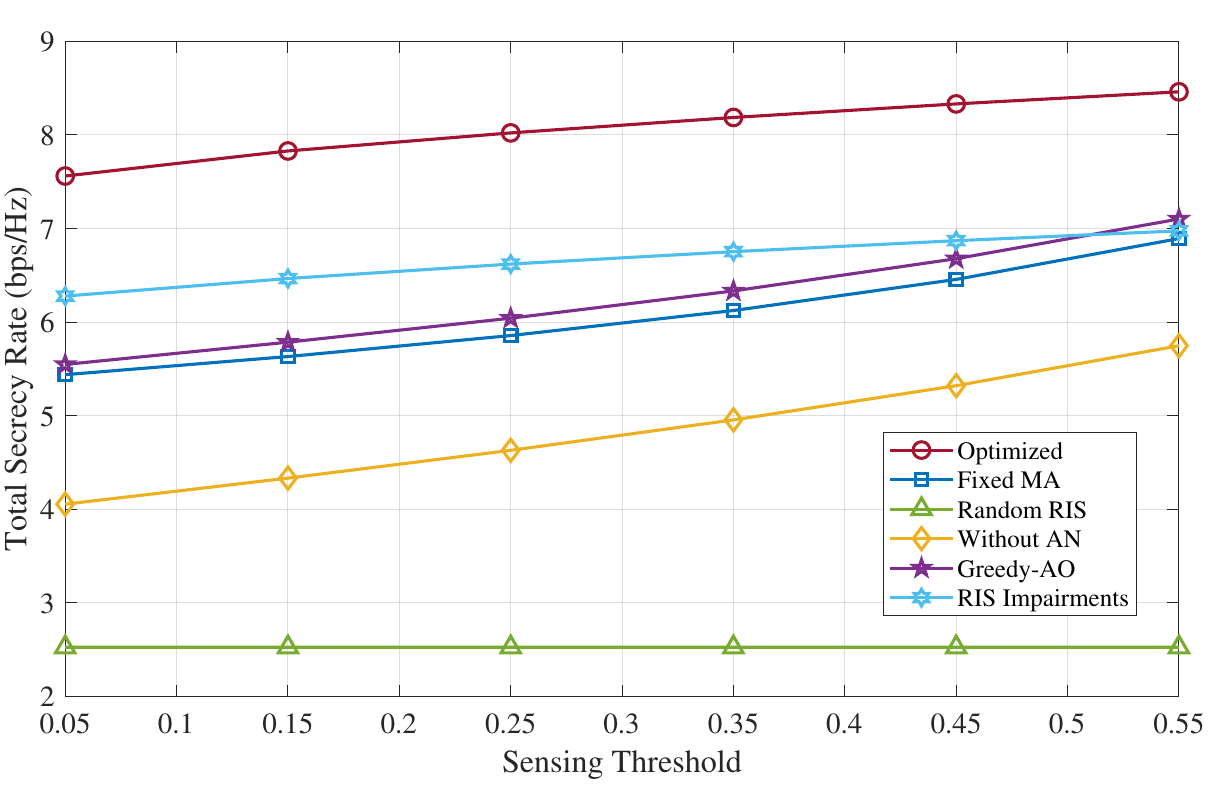} 
    \caption{\ Total secrecy rate versus sensing threshold.}
    \label{fig:7}
    \vspace{0.9em}
\end{figure}

Fig.~\ref{fig:7} illustrates the relationship between the sensing beampattern MSE threshold and the system's total secrecy rate. As the sensing threshold increases, the sensing constraint is relaxed, enlarging the feasible set of the joint optimization problem. Consequently, more spatial and power resources can be allocated to information beamforming, interference suppression, and AN design, leading to an improved secrecy rate. In contrast, the nearly unchanged random-RIS curve indicates that its performance is mainly limited by the unoptimized RIS phases rather than the sensing constraint. The gradual saturation of the RIS Impairment scheme further shows that, under a loose sensing requirement, hardware phase errors become a dominant performance bottleneck.

\vspace{-1.5em}
\vspace{-0.3em}

\begin{figure}[H]  
    \centering
\includegraphics[width=0.90\linewidth]{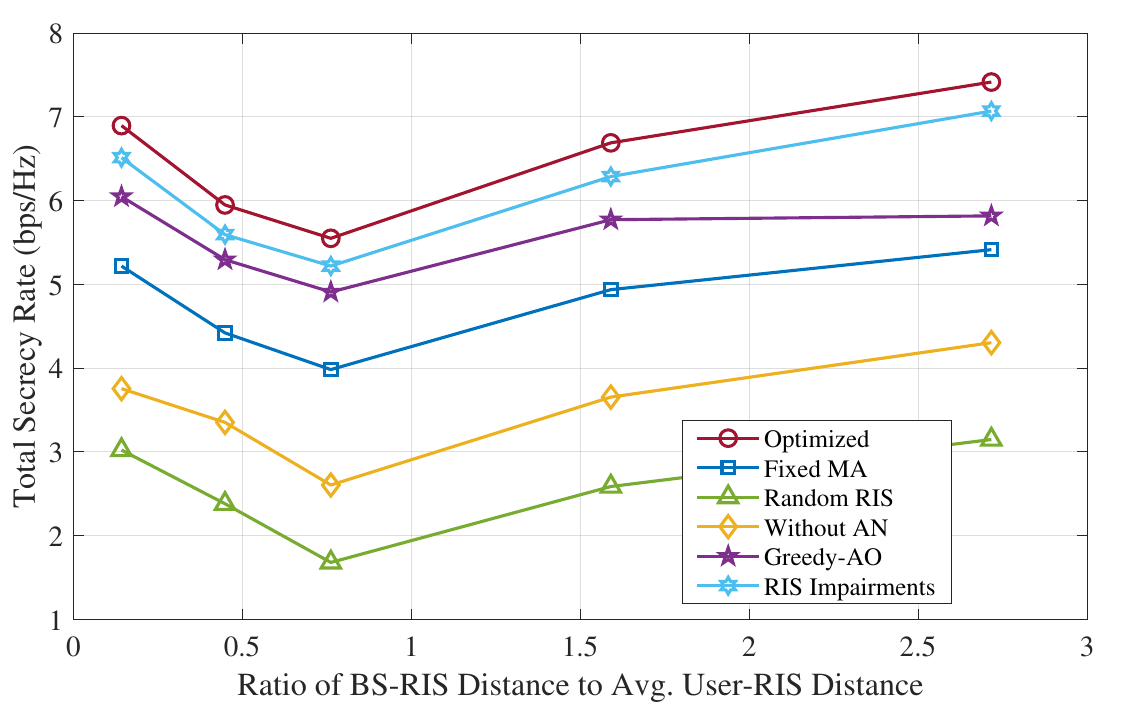} 
    \caption{\ Total secrecy rate versus the ratio of BS-RIS distance to avg. user-RIS distance.}
    \label{fig:8}
    \vspace{1.8em}
\end{figure}
\vspace{-0.3em}

Fig.~\ref{fig:8} illustrates the total secrecy rate versus the ratio of the BS-RIS distance to the average user-RIS distance. The secrecy rate initially decreases and then increases as the distance ratio grows. When the RIS is deployed close to either the BS or the legitimate users, the reduced path loss of the shorter hop partially compensates for the attenuation of the other hop. In contrast, at an intermediate distance ratio, both links experience considerable propagation loss, weakening the cascaded BS-RIS-user channels and reducing the available beamforming gain. The proposed scheme maintains the highest secrecy rate because the MA positions and active and passive beamforming variables are jointly adapted to the channel conditions. The improvement of Greedy-AO over Fixed MA further confirms that adaptive antenna positioning can mitigate the performance degradation caused by unfavorable RIS deployment.

Figs.~~\ref{fig:3}-~\ref{fig:8} evaluate the sensitivity of the proposed scheme to the number
of RIS elements, maximum transmit power, number of MA elements, number of
candidate positions, sensing threshold, and RIS deployment position under
the same user-location setting. Although the absolute secrecy rate varies
with these parameters, the proposed scheme consistently achieves the
highest performance over the considered ranges. 

\vspace{-1em}
\begin{figure}[H]  
    \centering
\includegraphics[width=0.92\linewidth]{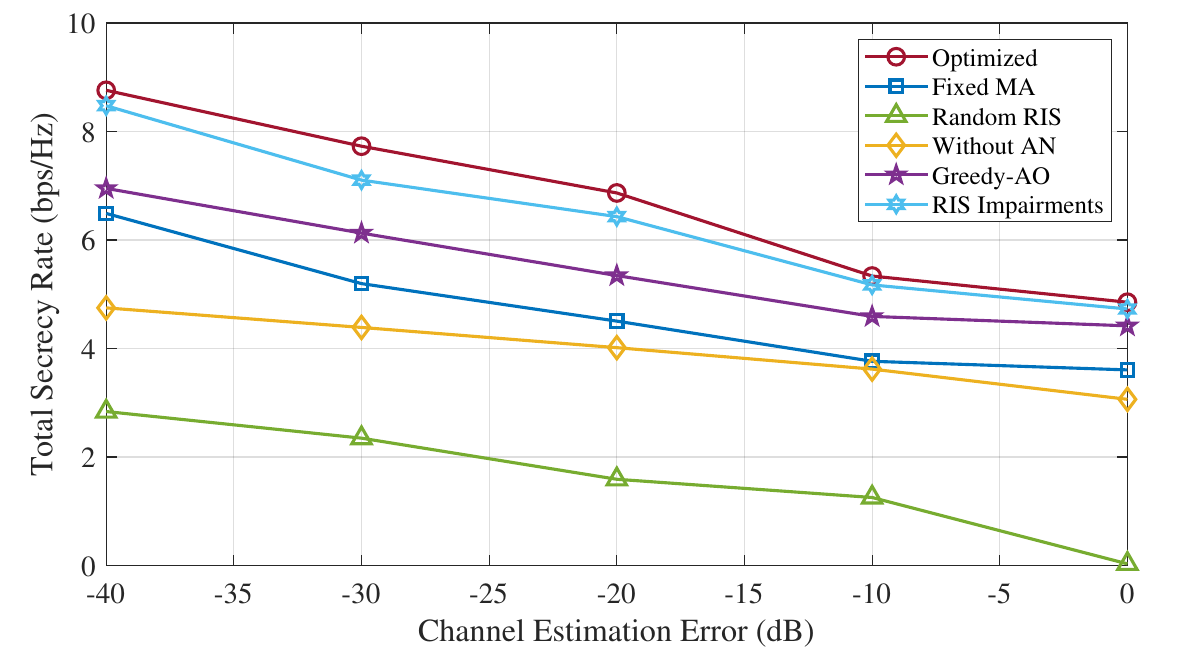} 
    \caption{\ Total secrecy rate versus the channel estimation error.}
    \label{fig:9}
    \vspace{2em}
\end{figure}
\vspace{-0.6em}

Fig.~\ref{fig:9} illustrates the total secrecy rate versus the channel estimation error (CEE), where a larger CEE represents less accurate CSI. As the CEE increases from -40 dB to 0 dB, the secrecy rates of all schemes decrease because the transmission variables optimized using the estimated channels become increasingly mismatched with the actual channels. This mismatch weakens coherent signal enhancement and interference suppression while reducing the accuracy of AN transmission toward the eavesdropper. Nevertheless, the proposed scheme consistently achieves the highest secrecy rate, demonstrating that the joint optimization of MA positions, active beamforming, AN covariance, and RIS phase shifts remains effective under imperfect CSI. The reduced performance gaps under severe channel estimation errors indicate that CSI uncertainty gradually becomes the dominant performance limitation.

\begin{center}

\begin{minipage}[t]{0.48\linewidth}
    \centering
    \includegraphics[width=\linewidth]{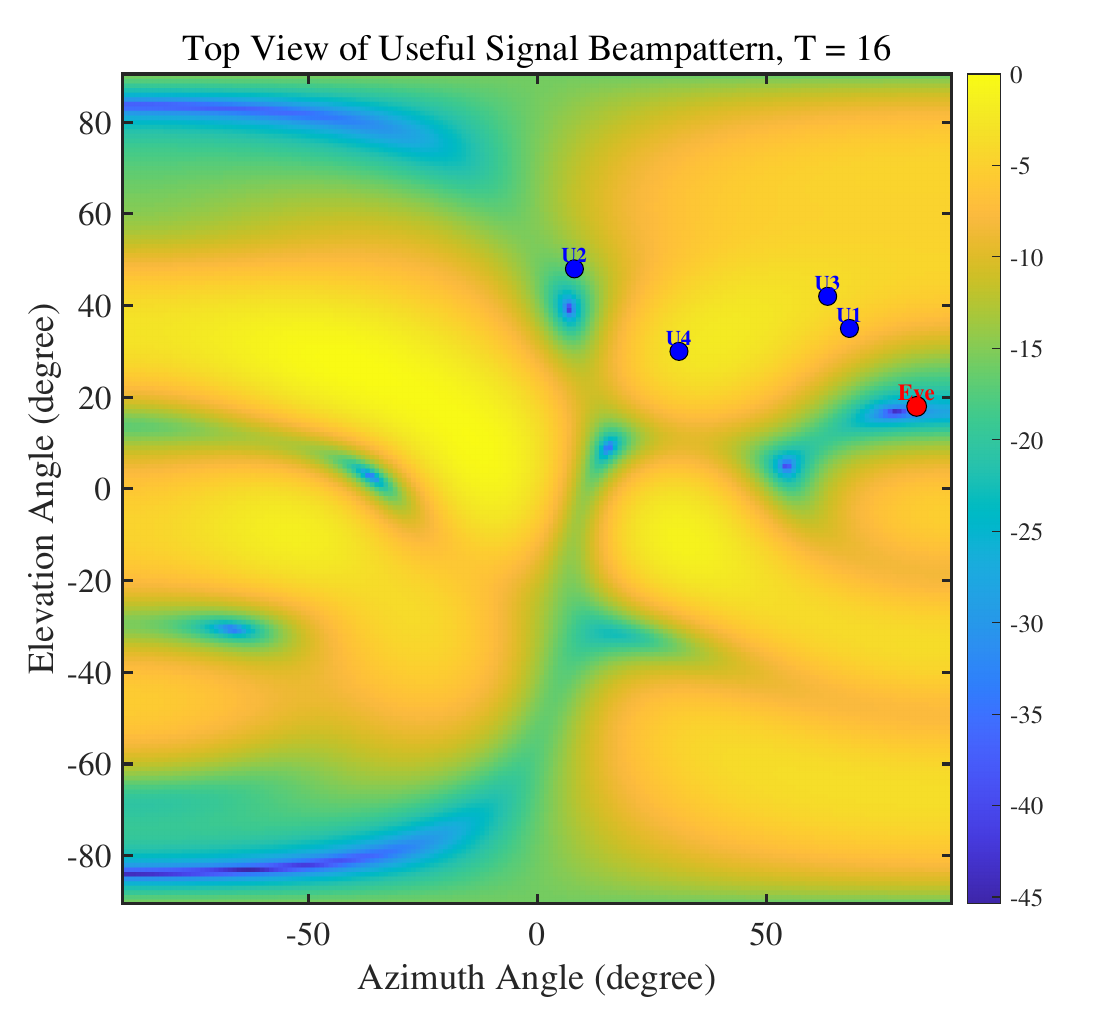}

    \vspace{0.1mm}
    {\footnotesize (a) Useful signal beampattern.}
\end{minipage}
\hfill
\begin{minipage}[t]{0.50\linewidth}
    \centering
    \includegraphics[width=\linewidth]{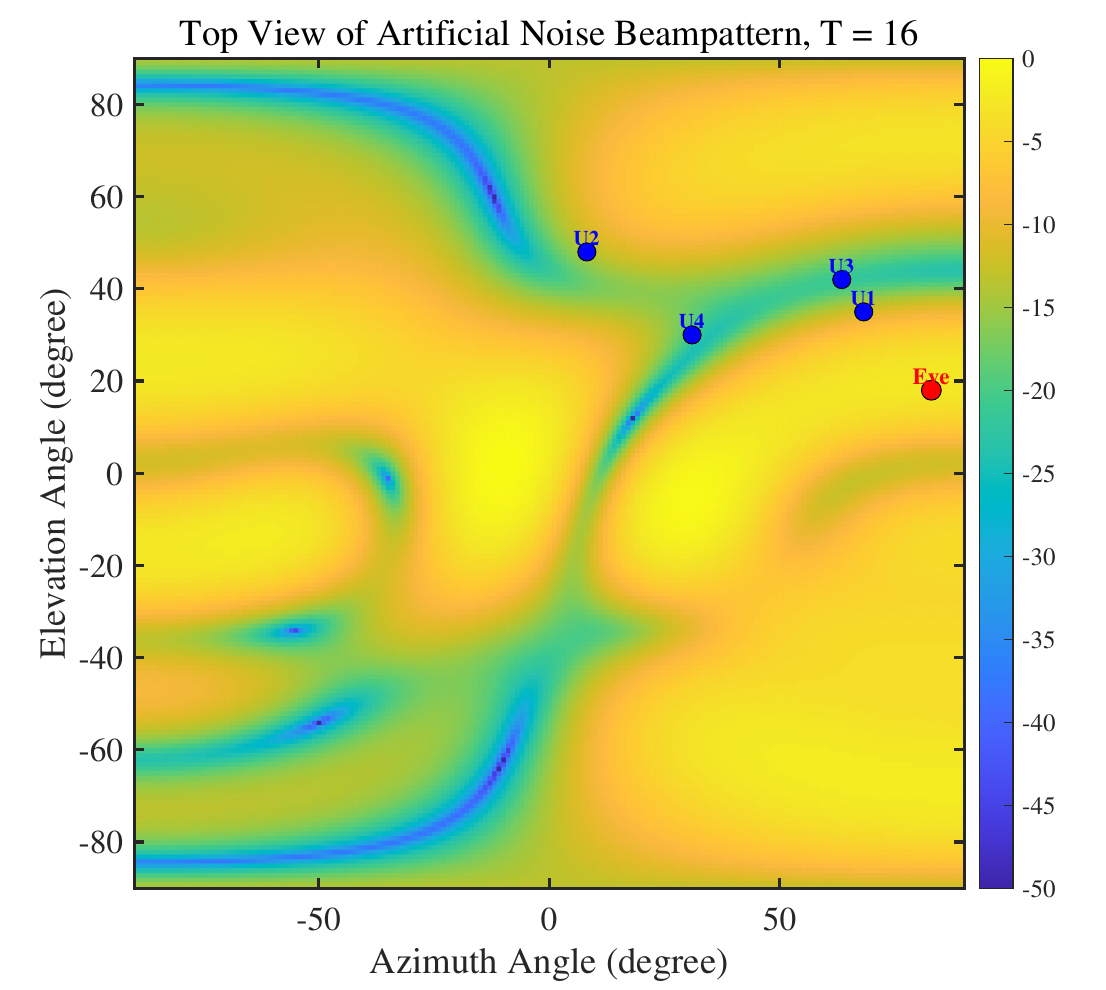}

    \vspace{0.5mm}
    {\footnotesize (b) AN beampattern.}
\end{minipage}

\captionof{figure}{Normalized beamforming visualizations of the proposed scheme.}
\label{fig:10}

\end{center}
\vspace{2em}

Fig.~\ref{fig:10} presents the normalized beampatterns of the useful signal and artificial noise. The useful signal beampattern enhances the spatial radiation toward the legitimate users, whereas the AN beampattern allocates stronger interference around the eavesdropper direction. These results provide an intuitive illustration of how the proposed joint optimization coordinates the MA positions, active beamforming, AN covariance, and RIS phase shifts to improve secrecy performance.

\vspace{-1em}

\section{Conclusion}\label{sec5}
This paper investigates discrete antenna positioning and beamforming design algorithms for RIS-assisted MA secure ISAC systems. First, we formulate a comprehensive system model for the RIS-assisted MA secure ISAC system. Based on this model, we develop a joint optimization problem with the objective of maximizing the system's total secrecy rate. To tackle the challenges posed by non-convex objective function and constraints, as well as the highly coupled optimization variables including MA position selection, beamforming design, AN and RIS phase shift matrices, we propose a joint optimization algorithm based on an AO framework. This approach incorporates various techniques including BPSO, SCA and DC programming to effectively handle constraints and optimize the variables iteratively. Finally, extensive numerical simulation experiments are conducted to compare the performance of the proposed algorithm against other baseline algorithms, validating its effectiveness and superiority in terms of security performance.
\vspace{-0.8em}

\balance
\bibliographystyle{IEEEtran}
\bibliography{reference}

\end{document}